\documentclass[preprint2]{aastex}
\usepackage{graphicx}

\begin{document}

\title{Spectral Analysis of the 13 keV Feature in XTE J1810-197: \\ Implications for AXP Models}

\author{N. Koning, R. Ouyed and D. Leahy}
\affil{Department of Physics and Astronomy, University of Calgary,
    Calgary, AB, Canada, T2N 1N4}

\begin{abstract}

During 2003 and 2004 the Anomalous X-Ray Pulsar XTE J1810-197 went through a series of four bursts.  The spectrum in the tail of one of these bursts shows a strong, significant emission feature ~13 keV, thereby encoding a wealth of information about the environment surrounding this object.  In this paper we analyse this emission feature considering both cyclotron and atomic emission processes and weigh our findings against three leading AXP models: the Magnetar model, Fall-back disk model and the Quark nova model.  We find that atomic emission from Rubidium within a Keplerian ring ($\sim$15 km from a compact object of $\sim 2M_\odot$) is the most consistent scenario with the observations, supporting the Quark nova model.  Cyclotron emission from an atmosphere a few hundred meters thick also fits the feature well, but is ruled out on account of its positional coincidence in three separate AXP sources.

\end{abstract}


\section{Introduction}

Anomalous X-Ray Pulsars (AXPs) are solitary objects characterized by persistent emission, X-ray pulsations and a rapid spin down rate. Rarely, some AXPs also emit Soft Gamma-ray Repeater (SGR) -like X-ray bursts. The first detected AXP was 1E 2259-586 by \citet{fahlman_1981}.  Since then, several more have been discovered including 1E 1048.1-5937, 4U 0142+61, 1E 1547.0-5408, 1RXS J170849.0-400910, 1E 1841-045, XTE J1810-197, CXOU J010043.1-721134, and CXO J164710.2-455216.  These objects exhibit spin periods of 5-12 seconds, fairly constant x-ray luminosities of $\sim 1\times10^{33}$ erg s$^{-1}$ to $\sim 1\times10^{36}$ erg s$^{-1}$, soft x-ray spectra and rapid spin down rates of $\dot{P}=1\times10^{-13} \to 1\times10^{-10}$ s s$^{-1}$ \citep{woods_2005}.  AXPs have some properties similar to SGRs, and are thought to be a related phenomenon.

The leading model to explain the AXP phenomenon is the Magnetar model \citep{duncan_1995}.  Magnetars are neutron stars (NS) with extreme magnetic fields between $\sim1\times10^{14}$ and $\sim1\times10^{15}$ G.  The bursting mechanism in the Magnetar model may be located on the NS surface, or in the magnetosphere.  In one model the large magnetic field exerts stress on the crust, causing it to fracture and subsequently emit short X-ray bursts.  These ``starquakes'' are believed to create additional hot-spots on the surface of the NS \citep{pons_2012}.  Another model suggests that reconnection events of a highly twisted magnetic field create the bursts in the magnetosphere \citep{lyutikov_2002}.  The spin-down rate and periods observed in AXPs are explained through a combination of magnetic-breaking and magnetic field decay on a time-scale shorter than the spin-down time-scale.

Another AXP model was put forth by \citet{chat_2000} in which a fossil accretion disk supplies the X-ray luminosity.  The disk is formed by fall-back material from the supernova (SN) explosion that created the NS.  The Fall-Back Disk (FBD) model requires modest magnetic fields ($B < \sim1\times10^{13}$ G, \citet{ertan_2008, ertan_2007}) with inner and outer radii of $\sim$10,000 km and $>$100,000 km respectively.  It should be noted that the FBD model does not explain the origin of the bursts and some hybrid magnetar models have been suggested \citep{eksi_2003, ertan_2003}.  Evidence of a fall-back disk in the source 4U 0142+61 has been reported by \citet{wang_2006}.  \citet{ertan_2006} provide further evidence of fall-back disks around other AXPs giving support to this theory.  This spin-down rates and periods observed in AXPs are explained through torques provided by the magnetic-field/disk interaction. \citep{ertan_2009}

In recent years a new theory to explain the properties of SGRs and AXPs has emerged using quark stars as the engine \citep{ouyed_2007a, ouyed_2007b}.  A quark star is created through the detonation of the neutron star in a process called a quark nova (QN) \citep{ouyed_2002, keranen_2005, niebergal_2010}.  The QN ejects the neutron-rich outermost layers of the neutron star \citep{ouyed_2009}, leading to r-process production of heavy elements \citep{jaikumar_2007}.  The QN can result in heavy-element-rich material from the ejected NS crust in a Keplerian ring orbiting the remnant quark star, extending from $\sim$20 km to $\sim$100 km depending on its age\citep{ouyed_2007b}.  This ring is distinguished from a fall-back disk, as it is much closer to the surface of the central object, and is made up of primarily degenerate material. The quark star is born with a dipole magnetic field aligned with its axis of rotation \citep{ouyed_2004, ouyed_2006} and a surface magnetic field of $B\sim1\times10^{15}$ G \citep{iwazaki_2005}.   In this model, X-ray bursts from AXPs occur when chunks of material break off the ring and accrete onto the quark star surface, while quiescent emission is powered by magnetic flux expulsion during spin-down.  In this model, the period and spin-down rate of AXPs is through dipole spin-down and magnetic field decay \citep{niebergal_2006}.

\subsection*{AXP Spectra}

During AXP quiescent phases, their spectra below $\sim$10 keV are often featureless with a continuum characterized by the sum of a power law and black body, the sum of two black bodies or even by resonant cyclotron scattering \citep{rea_2008}.  Broad\footnote{\citet{gavriil_2008} report ``very broad'' lines in all three sources in their analysis; \citet{woods_2005} report a $3\sigma$ upper limit of 4.1 keV width for XTE J1810-197. The different interpretations in line width is related to detector limitations, such as limited sensitivity and spectral resolution, and to different data analysis methods. We proceed here using the \citet{gavriil_2008} (Fig. 2) spectral data.} emission lines have been observed in the spectra of three AXPs during their bursting phase; 4U 0142+61, XTE J1810-197 and 1E 1048.1 \citep{gavriil_2008}.  In each case, there is a significant feature at $\sim$13 keV which has the potential to provide crucial information about the bursting environment.  In this paper we will concentrate on the spectrum from XTE J1810-197.

XTE J1810-197 was first observed in 2003 by \citet{ibrahim_2004}.  It is at a distance of $\sim$5 kpc \citep{gotthelf_2004}, has a period of 5.5 seconds and a spin down rate of $\dot{P}\sim1\times10^{-11}$ .  The inferred magnetic field due to magnetic braking is $\sim 3\times10^{14}$ G.  In 2005, XTE J1810-197 went through a series of four bursts \citep{woods_2005}.  During one of the burst tails, a significant emission line at $\sim$13 keV was detected.  Subtraction of the best-fit continuum model leaves a strong, significant emission feature at $\sim$13 keV (Figure 9 from \citet{woods_2005}).  \citet{gavriil_2008} re-analysed the data of \citet{woods_2005} in their comparison of spectral features from 4U 0142+61, XTE J1810-197 and 1E 1048.1.  Their re-analysed data is shown in Fig.\ref{fig:observed} with a linear energy axis. This work will focus on exploring what physical and geometrical conditions are required to reproduce the width and location of the 13 keV feature of XTE J1810-197 and the implications for AXP models.

In the Magnetar model, the line is commonly interpreted as a cyclotron line, implying field strengths $\sim 2.2 \times 10^{15}$ G for protons and $\sim 1.2 \times 10^{15}$ G for electrons \citep{gavriil_2008}.  In this work we argue that the large width of the 13 keV feature is naturally explained (see section \ref{section:constBField} below) by cyclotron radiation in a spatially varying magnetic field.  The width of the line then provides information on the extent of the emitting region in a dipole or multipolar field around the magnetar.  The cyclotron interpretation does have some challenges:  if due to protons, the inferred magnetic field ($\sim 2.2 \times 10^{15}$ G) is much larger than that derived by magnetic breaking ($\sim 3\times10^{14}$ G).  This can be reconciled with non-dipolar fields however, because the spin down rate depends on the dipolar component, whereas the line gives the field at the surface. If the line is an electron cyclotron line, the implied field is too low for the surface of the magnetar, and thus requires emission from far above the surface ($\sim$100 km due to the nature of a dipole field).  Another complication is that there are other significant lines detected (at $\sim$4 and $\sim$8 keV in 4U 0142+61).  If these are also proton cyclotron lines, then the emission must be coming from multiple, distinct regions on the magnetar surface,  possibly providing evidence for a multipolar field.  Perhaps the most damaging argument against the cyclotron interpretation of the 13 keV line is that it occurs in the three sources at the same energy,  implying that these sources all have an emission region with the same magnetic field. 

Another possible interpretation is that the 13 keV feature is an atomic line. This resolves the problem of why we see the same line in three different sources.  One of the challenges, as usual, is identifying which species is responsible for the line.  The high energy of the feature implies a heavy (heavier than Fe), highly ionized atom, which may be difficult to explain in some models.  Another challenge with the atomic line interpretation is the width of the feature. Doppler broadening requires velocities a good fraction of the speed of light which are not expected in or around a magnetar with a period of 2-10 s.

The purpose of this paper is to investigate the origin of the 13 keV feature by considering both cyclotron and atomic lines in different models.  The paper is organized as follows:  In section 2 we discuss the modelling procedure and the software we use.  Section 3 investigates the viability of the cyclotron interpretation by investigating the emission in several different scenarios.  Section 4 then examines the atomic line interpretation and its feasibility.  Section 5 discusses the results and the implications for the different AXP models.  Finally Section 6 provides our conclusions and predictions for future observations.

\section{Line Modelling with SHAPE}

In order to investigate both cyclotron and atomic emission from an astronomical source, we need software capable of performing radiative transfer.  Additionally, because we are exploring such a vast parameter space, we need the software to be fast and easy to use.  Software that meets these two requirements is SHAPE \citep{steffen_2011}\footnote{http://www.astrosen.unam.mx/shape/}.

SHAPE is a morpho-kinematic modelling tool developed to model astronomical objects and their environment in three dimensions (3D).  The environment within SHAPE allows one to easily construct 3D structures with arbitrary density, temperature and velocity profiles.  SHAPE then calculates the resulting emission and displays output in the form of two dimensional (2D) images, PV diagrams and spectra for comparison with observed data.

In order to keep the processing time at a minimum, SHAPE uses a ray-tracing radiative transfer algorithm.  The difference between this approach and other radiative transfer code is that SHAPE requires certain attributes (e.g. temperature, density, velocity, emitting species etc...) be defined as input, instead of them being calculated by the code.  SHAPE therefore answers the question ``Given a temperature, density and velocity distribution, what is the resulting spectrum from a particular species?'' 

It should be noted that SHAPE does not take radiative effects specific to high magnetic fields into account.  For example, birefringence, photon splitting, scattering suppression and distortion of atoms in large magnetic fields are not taken into account.  

\subsection{Algorithm and Spectral Modelling}

SHAPE partitions 3D space into an axis aligned (with the observer) grid of 256$^3$  cells. Each cell within the volume of a structure (e.g. sphere, torus, disk) is assigned attributes (e.g. temperature, density, velocity, pressure, magnetic field, gravitational field) from which emission and absorption coefficients are calculated depending on the species involved (e.g. atoms, molecules, electrons, protons).  A ray is then passed through the grid for each pixel of the detector (computer screen) travelling through multiple cells where radiation is either added to, or removed from the ray.  Each ray consists of $m$ frequency bands evenly spaced between a minimum and maximum frequency range, set to match the observed data.  The energy within each band of each ray can then be plotted, generating a final spectrum that can be compared with observations.  For more information on the algorithms involved in SHAPE, see http://www.astrosen.unam.mx/shape/.

For the cyclotron simulations, SHAPE assigns each cell in the grid a magnetic field based on the selected field type (dipole or constant) and its location in space.  The field within each cell is therefore taken as constant and we do not consider the effects of a charged particle travelling from one cell to another.

For each of the models presented in this work, spectra will be produced with SHAPE in the aforementioned manner.  The energy range will span $\sim$3 - 30 keV in order to match the observed spectra of XTE J1810-197 and divided into 100 frequency bands.  When comparing our best-fit models with the data, we apply a Gaussian convolution with FWHM of 1.5 keV to simulate the energy resolution of the observations.  A grid of 256$^3$ will be used in all models.  

\section{Cyclotron Emission}

Cyclotron radiation is the result of a moving charged particle being deflected in a static magnetic field.  Due to the Lorentz force, a moving charged particle will undergo circular motion perpendicular to its velocity and the magnetic field.  For particles that are moving only slightly relativistically, radiation is primarily emitted at the cyclotron frequency:

\begin{equation}
\omega_c = eB/\gamma m
\label{eq:omegac}
\end{equation}

Where $e$ is the charge of the particle, $B$ is the magnitude of the magnetic field, $\gamma$ is the relativistic correction and $m$ is the mass of the charged particle. Lines can also be emitted at harmonics of the fundamental frequency, their line centres given by $\omega_l=l\times\omega_c$, where $l$ is an integer ranging from $l \to \infty$.  The line center of the emission is therefore proportional to the magnitude of the magnetic field. The emission coefficients for the cyclotron lines are given by \citet{boyd_2003}:

\begin{eqnarray}
j_l(\omega,\theta)&=&\frac{n e^2 l^2 \omega _c^2 l^{2l}}{16 \pi^2 \epsilon _0 c [l!]^2}(\cos ^2(\theta)+1)\sin^{2(l-1)}(\theta) \nonumber \\
&\times& \left(\frac{k_B T_e}{2 m c^2}\right)^l \left(\frac{m c^2}{2 \pi k_B T_e} \right)^{1/2}
(\omega _l \cos(\theta))^{-1} \nonumber \\
&\times& \exp \left[\frac{-m c^2 (\omega - \omega _l)^2}{2 k_B T_e \omega _l^2
\cos^2(\theta)}\right]
\end{eqnarray}

Where $n$ is the density of the charged particles, $l$ is the harmonic number, $\theta$ is the angle between the magnetic field and the observer, $k_B$ is the Boltzmann constant, $T_e$ is the electron temperature, and $m$ is the mass of the charged particle.  In the regime where $k_B T_e / 2 m c^2 \ll 1$, the fundamental frequency will dominate the spectrum.  For this reason we will concentrate on the fundamental line, $l=1$.  The emission coefficient then takes the simpler form:

\begin{eqnarray}
j(\omega,\theta)&=&\frac{n e^2 \omega _c^2 k_B T_e}{32 m \pi^2 \epsilon _0 c^2}
\left(\frac{\cos ^2(\theta)+1}{\omega _c \cos(\theta)}\right) 
\left(\frac{m}{2 \pi k_B T_e}\right)^{1/2} \nonumber \\
&\times&  \exp \left[\frac{-m c^2 (\omega - \omega _c)^2}{2 k_B T_e \omega _c^2\cos^2(\theta)}\right]
\label{eqn:jsimple}
\end{eqnarray}

Equation \ref{eqn:jsimple} represents a Gaussian line profile with amplitude:

\begin{equation}
a = \frac{n e^2 \omega _c^2 k_B T_e}{32 m \pi^2 \epsilon _0 c^2}
\left(\frac{\cos ^2(\theta)+1}{\omega _c \cos(\theta)}\right) 
\left(\frac{m}{2 \pi k_B T_e}\right)^{1/2}
\label{eq:amplitude}
\end{equation}

and width:

\begin{equation}
\Delta \omega_{\textsf{FWHM}} \approx \left[\frac{2.35 k_B T_e \omega_c^2 \cos^2(\theta)}{m c^2}\right]^{1/2}
\label{eq:dOmega}
\end{equation}

For a given temperature and density, the intensity of the line is proportional to the angle of the observer and the magnetic field, and the mass of the charged particle:

\begin{equation}
a \propto \frac{B}{m^{5/2} }\left[\frac{\cos^2(\theta) + 1}{\cos(\theta)}\right]
\end{equation}

This implies that as the angle increases, so will the intensity and a proton is expected to produce a less intense line than an electron.  The width of the line is also proportional to the angle between the observer and the magnetic field and the mass of the charged particle:

\begin{equation}
\Delta \omega_{\textsf{FWHM}} \propto \frac{B \cos(\theta)}{m^{5/2}}
\end{equation}

Smaller angles will create a broader line, and electrons will produce broader lines than protons.  For a given line center, $\omega_c$ , we can expect the electron line to be $\approx 43 \times$ broader than that produced by protons.

We wish to investigate the cyclotron line shape for both electrons and protons in a variety of environments in an effort to resolve the origin of the 13 keV emission feature seen in XTE J1810-197.

\subsection{Constant Magnetic Field}

We begin by investigating the trivial case of thermalized electrons and protons in a constant magnetic field. The geometry of the model is that of a spherical shell with an inner and outer radius of 10 km and 12 km respectively.  The density within the shell is constant, as is the temperature which is set to $1\times10^8$ K. 

We model both proton and electron cyclotron radiation with a constant magnetic field of $2.15\times10^{15}$ G and $1.15\times10^{15}$ G respectively.  We use these field strengths so that the line centres match up with the 13 keV feature in the spectrum of XTE J1810-197.  Since the line shape depends on the angle between the magnetic field and the observer, we will examine two scenarios; one with the magnetic field at $30^{\circ}$ to the observer and one at $60^{\circ}$.

The left panel of Fig.\ref{fig:ep_cylo_Bconstant} shows both the electron and proton results. The most dramatic result between the proton and electron lines is their relative breadth.  The breadth in the proton case, at both $30^{\circ}$ and  $60^{\circ}$, does not exceed one frequency band in our simulation.  On the other hand, the electron cyclotron lines are much broader.  This is of course due to the mass dependency of the line width (see equation \ref{eq:dOmega}).  The electron line under the $30^{\circ}$ inclination is more intense but narrower than the  $60^{\circ}$ line.  This again can be predicted by equations \ref{eq:amplitude} and \ref{eq:dOmega}.  The width of a cyclotron line is therefore heavily dependent on both the mass of the particle and on the angle between the observer and the magnetic field.

From this simulation we can see that the breadth of the 13 keV feature can easily be obtained through electron cyclotron emission in a constant magnetic field.  For proton cyclotron emission, the larger mass of the proton severely limits the line breadth and another mechanism is required for its broadening.

\subsection{Constant Magnetic Field + Variation}
\label{section:constBField}

A perfectly uniform, constant magnetic field is not realistic and some variation should be expected.  The next simulation investigates the shape of the proton cyclotron line in a non-uniform, variable magnetic field.  For this test, the magnitude of the magnetic field at each location within the shell will be randomly chosen between $2.15\times10^15$ G $\pm dB$ , and oriented at $60^{\circ}$ to the observer.  We will run two simulations, one with $dB=0.5\times10^{15}$ G and one with $dB=1.0\times10^{15}$ G.

The right panel of Fig.\ref{fig:ep_cylo_Bconstant} shows the results of this second simulation.  As expected, as the variation in the magnetic field ($dB$) increases, the breadth of the line increases.  A variation in the magnetic field of $dB=1.0\times10^{15}$ G is sufficient at $60^{\circ}$ for proton cyclotron radiation to reproduce the width of the 13 keV feature.

\subsection{Dipole Magnetic Field}

The previous simulations deal with the idealized case of charged particles in a (relatively) constant magnetic field.  In realistic situations, such as the space around a magnetar, the magnetic field configurations are more complex.  It is tempting to conclude that because the cyclotron line center is proportional to the magnetic field, a continuous, variable magnetic field like that of a magnetic dipole will produce a flat continuum of radiation, and not a distinct line.  The situation is complicated, however, by the dependence of the line strength on the angle between the field and the observer.  The next set of simulations is meant to investigate this situation by studying the cyclotron line properties produced from a spherical shell within a magnetic dipole field.

The parameters of this simulation are the same as before; a spherical shell with an inner and outer radius of 10 km and 12 km respectively and a constant temperature of $1\times10^8$ K.  The angle between the magnetic moment of the dipole and the observer is fixed at $90^{\circ}$.  For the proton cyclotron case, the magnetic dipole moment is $\mu_m = 2.0 \times 10^{30}\texttt{ A m$^2$}$ such that the magnetic field inside the spherical shell is $\sim 3 \times 10^{15}$ G and within our spectral range.  For the electron cyclotron case, we use $\mu_m = 1.0 \times 10^{27}\texttt{ A m$^2$}$ .  To show the dependency of the line shape on the size of the emission region, we have plotted the emission from a shell 1 km and 2 km in thickness, with the outer radius in both cases held at 12 km.

The left panel of Fig.\ref{fig:ep_cylo_shell} shows the results for the electron case and the right panel the results for the proton case.  The dipole field introduces several levels of complexity over the constant magnetic field case which is represented by the spectra.  Each point within the spherical shell has a magnetic field with a different direction and magnitude.  Since the cyclotron lines are highly sensitive to the angle between the field and the observer, the location and intensity of the lines from each location in space can be highly variable.  We can see this variability most clearly in the proton cyclotron spectrum in the form of``spiky''\footnote{The spectrum shows these spiky features due to the finite size of the individual grid cells in our simulation} features.  Interestingly, however, the sum of each individual line contributes to a broad, asymmetrical feature and not a flat continuum.  The electron case, as expected from the greater width of each individual line, shows a broader feature with less variability. In both the electron and proton cases, the thinner shell model resolves two distinct features, whereas the thicker shell blends the features into one broad one.  These broad features, especially when binned to instrument resolutions, may explain one or more features in the AXP spectra.   

\subsection{Dipole Magnetic Field – Polar Regions}

It is entirely likely that the emission of the cyclotron lines does not come from a complete spherical shell, but a small region perhaps around the poles of a dipole or multipole field.  This next simulation examines the consequence of the cyclotron lines emitted in a region of space near the poles of the dipole field.  We use the same configuration as before, except the spherical shell is limited to a region between $\phi=0^\circ$ and  $\phi=20^\circ$ where $\phi$ is the angle measured from the pole ($\phi=0^\circ$) to the equator ($\phi=90^\circ$).  As with the previous simulation, we simulate both a thicker shell (2 km) and a thinner shell (1 km).  The results are shown in the left and right panels of Fig.\ref{fig:ep_cylo_pole} for the electron and proton cases respectively.  We can clearly see that in both cases, the thickness of the shell contributes to the thickness of the emission feature.  The width of the observed 13 keV feature can therefore be matched in both the electron and proton cases with emission from a thin shell around the polar region of a dipolar magnetic field.

\subsection{Dipole Magnetic Field – Polar Regions - Rotated}

The 13 keV feature is fit quite well by either proton or cyclotron emission from a small region around the poles of a magnetic dipole field.  We are aware, from our previous simulations, of the effect of field orientation on the intensity and width of the line.  This next simulation therefore builds on the previous by examining the consequences of rotating the dipole field with respect to the observer.  We run four simulations, with the magnetic moment at $90^{\circ}$, $60^{\circ}$, $30^{\circ}$ and $0^{\circ}$ degrees to the observer.  Fig.\ref{fig:ep_cylo_poleangles} show the results for both the electron and proton cases.  As the angle between the dipole moment and the observer decreases, the line appears to shift to the red.  Interestingly it does not get much broader, but roughly keeps its original shape.  This may seem counter intuitive since the position of the line, according to equation \ref{eq:omegac}, depends only on the magnitude of the magnetic field.  This is true in a constant field, but in this situation we have a range of fields so we have emission over many frequencies.  It is the orientation that determines which lines are enhanced and suppressed.  At steeper angles, the polar regions are enhanced (which also have the strongest magnetic field and highest energy lines).  As the angles decrease, the polar regions are suppressed and the surrounding regions, which have lower magnetic fields, are enhanced resulting in an emission feature at lower energies.

\subsection{Conclusion}

The above simulations were meant to show how the cyclotron emission line looks under various conditions.  We have shown that the width and intensity of the line depends heavily on both the mass of the charged particle and the angle between the magnetic field and the observer.  Further adding to the breadth of the line is the variation in the magnetic field.  Looking at the realistic case of a polar emission region in a dipolar field, we see that the breadth of the line depends more on the thickness of the emission region than the mass of the particle.  Somewhat surprisingly, the direction of the magnetic dipole moment with respect to the observer along with its magnitude determines the center of the emission feature.  

We can now put all of this information together and try to fit the 13 keV feature seen in XTE J1810-197.  Fig.\ref{fig:ep_cylo_bestfit} shows the best fit models for the electron and proton cases in the left and right panels respectively.  In both cases we have a dipole magnetic field and the emitting region is a spherical shell restricted to the angles $\phi = 0^\circ \to 20^\circ$ .  In both cases, we require a thin shell with an inner and outer radius of 11.9 km and 12 km respectively.  For the electron case, the magnetic dipole moment is $\mu_m = 1.0 \times 10^{27}\texttt{ A m$^2$}$ inclined at $66^\circ$ to the observer.  For the proton case, the magnetic dipole moment is $\mu_m = 2.0 \times 10^{30}\texttt{ A m$^2$}$, inclined at $56^\circ$ to the observer.

Of course these fits are not unique as the same line profile can be achieved in several ways.   For example, the line center can be shifted by changing the angle of the dipole field with respect to the observer, or simply adjusting the magnitude of the magnetic dipole moment.  What these fits do tell us is that both electron and proton cyclotron emission can explain the 13 keV feature seen in XTE J1810-197 under reasonable physical and geometrical conditions.

\section{Atomic Line Emission}

In this section we will investigate the idea that the 13 keV feature in the spectrum of XTE J1810-197 may be caused by atomic line emission.  There are several questions concerning the atomic line interpretation.  First, is there a viable atomic transition near 13 keV and second, how do we explain the large breadth of the 13 keV feature.  By fitting the 13 keV feature with a Gaussian profile, we find that the center of the line is at 13.07 keV, the standard deviation is $\sigma \approx1.45$ keV, and the full-width-half-max is $\Delta E_{1/2} \approx 3.41$ keV.

We use the National Institute of Standards and Technology (NIST) atomic line database to identify possible counterpart lines to the 13 keV feature of XTE J1810-197.  Table \ref{table:transitions} lists all the available transitions between $\sim$9 keV and $\sim$18 keV with determined transition probabilities.

The most likely species for the counterpart line is Rubidium (Rb).  We come to this conclusion for several reasons.  First, and most important, is that only Rb lines have sufficient energy to be a viable candidate.  There are four Rb transitions near 13 keV: 14.3 keV (RbXXXVII), 14.2 keV (RbXXXVII), 13.9 keV (RbXXVI) and 13.8 keV (RbXXVI).  The next closest candidate is Ni with an energy of 10.356 keV, which is much too low.  Note that all the Rb lines are actually above 13.07 keV.  This is not a problem since when these lines are gravitationally red-shifted, they will fall closer to the observed feature.  Second, the Rb lines at $\sim$13.8 keV all have the highest transition probabilities of any line in the entire spectrum (lines spanning the whole spectrum were not listed in the table due to the large number of transitions).  This and the fact that several of these lines are extremely close, makes a strong feature at $\sim$13 keV very likely. 

Fig.\ref{fig:atomic_identification} shows the line identifications of the transitions listed in Table \ref{table:transitions} plotted with the observed spectrum from XTE J1810-197.  It is clear from this figure that Rb is the most likely candidate for the 13 keV feature.  Interestingly several weaker Rb transitions lie close to the energy of a small ``bump'' around 17 keV. This bump, however, may or may not be real.

Two ionization states of Rb may contribute to the observed feature; RbXXXVI and RbXXXVII.  The ionization energy of RbXXXV (giving RbXXXVI) is 4.36 keV and that of RbXXXVI (giving RbXXXVII) is 18.31 keV \citep{sansonetti_2006}.  Using $E=kT$ we find that this translates to temperatures on the order of $T \approx 5.0\times10^7$ K and $T \approx 2.0 \times10^8$ K for RbXXXVI and RbXXXVII respectively.  For this study we will assume a temperature of $T=1.5\times10^8$ K. We are unsure of the electron density, so we assume an abundance of RbXXXVI 10\% that of RbXXXVII.

\subsection{Line Modelling}

We will model the emission from Rb assuming it is optically thin and that we are in LTE.  The emission coefficients for a bound-bound transition from level $i \to j$ are calculated via:

\begin{equation}
j_\nu = \frac{h\nu}{4\pi}n_i A_{ij}
\label{eq:jbb}
\end{equation}

Where $n_i$ is the number density of atoms in the $i$th state and $A_{ij}$ is the transition probability.  In LTE we can use the Boltzmann equation to determine the number density for state $i$ given a temperature $T$ and total density $n$.  Employing equation \ref{eq:jbb} for the Rb transitions in \ref{table:transitions}, we can use SHAPE to produce a model spectrum. The left panel of Fig.\ref{fig:atomic_rb_lines} shows the resulting spectrum for a temperature of $T=1.5 \times 10^8$ K and a resolution of 1000 frequency bands in order to capture the finest structure.

\subsection{Gravitational Red-shift}

With the most likely species selected, we can now go ahead and investigate the physical and geometric conditions required for the Rb line to reproduce the 13 keV feature seen in XTE J1810-197.  We will use the Rb XXXVII transition at 14.299 keV as our reference line (most intense line in the left panel of Fig.\ref{fig:atomic_rb_lines}).  

The gravitational red-shift near a massive object is given by:

\begin{equation}
E_{o} = E_{e}\sqrt{1-\frac{2GM}{c^2r}}
\label{eq:gravshift}
\end{equation}

Where $E_{o}$ is the observed energy, $E_{e}$ is the emitted energy, $G$ is the gravitational constant, $M$ is the mass of the compact object and $r$ is the distance from the object. Solving for $r$, we can determine the location of the emission above the compact object:

\begin{equation}
r = \frac{2GM}{c^2}\left[\frac{E_{e}^2}{E_{e}^2-E_{o}^2}\right]
\label{eq:rgrav}
\end{equation}

For a mass of $1.4M_\odot$, and a shift from  $14.299 \to 13.07$ keV, we get an emitting radius of 25 km from the center of the compact object.  If we instead take the closer line, Rb XXXVI at 13.794 keV, we get an emitting radius of 60 km.  If the 13 keV feature in XTE J1810-197 is indeed an Rb atomic line, it must therefore be emitted from at least 25 km from the center of a compact object of $1.4M_\odot$ (assuming relatively small velocities so relativistic beaming is not a factor).

We model this scenario by placing a spherical shell with an inner and outer radius of 24.5 km and 25.5 km respectively in a gravitational field produced by a  compact object at the origin.  The right panel of Fig.\ref{fig:atomic_rb_lines} shows the resulting spectrum, the strongest line matching perfectly with the center of the 13 keV feature.

\subsection{Line Broadening}

Next we will look at what conditions can lead to an atomic line as broad as that seen in XTE J1810-197.  The two most likely broadening mechanisms for the 13 keV feature are Doppler broadening due to the bulk motion of the atoms, and thermal broadening due to their random motion.  Using the relativistic Doppler formulation, the line-of-sight velocity corresponding to a shift in energy $E_{e} - E_{o}$ is:

\begin{equation}
v = c\frac{(E_{e}^2-E_{o}^2)}{(E_{e}^2+E_{o}^2)}
\label{eq:vdoppler}
\end{equation}

At the $2 \sigma$ level, $E_o=15.97$ keV, we have velocities of $v_{2 \sigma} \approx 0.20c$ and at the $3 \sigma$ level the velocities reach as high as $v_{3 \sigma} \approx 0.29c$.  If the width of the line is due to non-relativistic, Maxwellian thermal broadening, this translates into a temperature of \citep{kwok_2007}:

\begin{equation}
T = m\left[\frac{c^2 \sigma ^2}{k_B E_{e}^2}\right]
\label{eq:Tbroad}
\end{equation}

The lowest possible temperature (using hydrogen for the mass) is therefore $T \approx 1.1 \times 10^{11}$ K. For Rubidium we require a temperature of $T \approx 9.5 \times 10^{12}$ K.  These temperatures are extreme and not expected in any AXP model.  We will therefore not consider this option further.

Since these speeds are a significant fraction of the speed of light, relativistic beaming is expected to be important and may alter the line shape.  The intensity at a specific frequency is scaled as:

\begin{equation}
I_o = D^3I_e
\label{eq:relbeam}
\end{equation}

Where $I_o$ and $I_e$ are the observed and emitted intensities respectively,$D=1/\gamma(1-\beta\cos(\theta))$ is the Doppler factor, and $\theta$ is the angle between the observer and the particle velocity.  Due to relativistic beaming, we expect to see the blue emission increased, and the red suppressed assuming motion toward/away from the observer.  This should also cause the peak of the line to shift towards the blue.

There is another possible broadening mechanism that can be significant near a massive, compact object.  Near such an object, the gravitational field may significantly shift the line center towards the red as given by equation \ref{eq:gravshift}.  If the emission region is extended, however, the material closer to the mass will experience a greater red-shift than the material further away.  The net effect will be ``Gravitational Broadening'' of the spectral line with the breadth conveying the extent of the emission region.

With the above considerations, we will now model these different broadening mechanisms in an effort to determine under what conditions we can get a broad enough line using atomic transitions.

\subsubsection{Doppler Broadening}

We will investigate Doppler broadening of the Rb line under two different geometries: a spherical shell and extended disk.  The velocity profile will be one of rotation, where the velocity direction is perpendicular to the y-axis and the position vector (observer line-of-sight is along the z axis).  The magnitude of the velocity will be constant at 0.2c.  The density is constant, as is the temperature which is set to $T=1.5\times10^8$ K.  We use 100 frequency bands.

\subsubsection*{Spherical Shell}

The first model we investigate is a spherical shell with an inner and outer radius of 24.5 km and 25.5 km respectively.   The left panel of Fig.\ref{fig:atomic_rotation} shows the results of this simulation.  We incline the sphere (rotational plane) at $90^\circ$ and $30^\circ$ to the observer to demonstrate the effects of inclination on the spectrum, with $0^\circ$ being pole-on.  As predicted, the width of the lines roughly match that of the 13 keV feature and the Doppler broadening suppresses the red while enhancing the blue.  In the $90^\circ$ rotation case, this leads to a shift of the line center towards the blue.  The $30^\circ$ rotation of the sphere means the line of sight velocity is decreased, and we therefore see less of a shift.  Because of the relativistic beaming, the center of the line in the $90^\circ$ rotation case is shifted towards higher energies.  In order to match the 13 keV feature, we require a larger gravitational red-shift which means the emitting region must be closer than the 25 km calculated above.

\subsubsection*{Disk\footnote{A ``disk'' has different meanings for each of the AXP models.  In the QN model, the disk is made up heavy-element-rich degenerate material and located $\sim$20 km to $\sim$100 km away from the quark star.  The disk of the FBD model is formed from the debris of the supernova.  It is much larger and located further away from the central star.}}

The next model we investigate is a rotating disk.  The disk has an inner radius of 24.5 km, an outer radius of 25.5 km and a height of 5 km.  We model the system at two inclinations: $90^\circ$ and $30^\circ$, with the results displayed in the right panel of Fig.\ref{fig:atomic_rotation}.  The resulting line is broadest at an inclination of $90^\circ$ and narrowest at $30^\circ$.  This is expected since the line of sight velocity is maximum when the inclination of the disk is $90^\circ$.  It is also apparent that the feature is being split at the lower inclinations.  This is a common spectral profile of a rotating disk.  The red-shifted line is lower than the blue-shifted because of relativistic beaming.

\subsubsection{Gravitational Broadening}

If emission is from an extended atmosphere in a large gravitational field, it is possible for a spectral line to experience gravitational broadening. Using equation \ref{eq:rgrav} we can get an idea of the extent of the atmosphere.

Immediately we run into problems with this analysis.  The Rb line is emitted at 14.299 keV but the feature is observed to extend blue-ward of this.  Gravitational red-shift only shifts the line to the red, making this scenario impossible (again assuming relativistic beaming is negligible).  However, there are weaker Rb lines at greater energies ($\sim 16 \to \sim 19$ keV) which may be contributing to the extended breadth of the line.  For this reason we will consider the maximum possible energy of the line under gravitational broadening to be 14.299 keV.  Under this assumption, the line therefore extends from $E_{\texttt{red}} \sim 11.84 \to E_{\texttt{blue}}=14.299$ keV. For a compact object of $1.4M_\odot$  this translates, through equation \ref{eq:rgrav}, to an inner radius of $r_{\texttt{in}} \sim 13$ km.  The outer radius, in theory, must extend to infinity.  If we take a more reasonable maximum energy of the line to be $E_{\texttt{blue}} \sim 14.2$ keV we find the outer radius must be at least $r_{\texttt{in}} \sim 300$ km.

We notice, by equation \ref{eq:gravshift}, that most of the energy of the resulting feature will be emitted closer to the non-shifted energy (more emitting atoms are at larger radii).  For the red side of the feature to be observable, a density gradient is therefore required.  We model this situation in SHAPE by using a disk with an inner and outer radius of 10 km and 300 km respectively in a gravitational field of a $2M_\odot$ compact object (we use $2M_\odot$ instead of $1.4M_\odot$ to make the effect more pronounced)   We model three different density distributions, the results of which are shown in Fig.\ref{fig:atomic_gravbroad}.  Panels a, b and c show the line profile using $n=\texttt{constant}$, $n \propto 1/r^2$ and $n \propto 1/r^3$ respectively.  We can see that as the density profile becomes steeper, the line becomes broader.  This is because there is more of the inner material where the largest red-shift is occurring.  Panel d of Fig.\ref{fig:atomic_gravbroad} shows the results of the $n \propto 1/r^3$ profile but convolved with a Gaussian with FWHM of 1.5 keV to better match the observations.  We can see that the breadth and location of the line can be reproduced in this scenario.

\subsection{Conclusion}

The 13 keV feature of XTE J1810-197 may be explained by line emission from Rb XXXVI or RbXXXVII.  If the feature is indeed from an atomic line, the atoms must be either moving at mildly relativistic speeds, emitted from an extended atmosphere in a large gravitational field, or perhaps a combination of both.  Although the feature is probably a blend of many Rb lines, the 14.299 Rb XXXVII line is the most likely candidate for line center.  In order to explain the offset in energy between this Rb line and the observed feature, the emission must be coming from a region at least 25 km from the origin assuming a compact object of $1.4M_\odot$.  We have seen, however, that when large velocities (0.2c in this case) are involved, relativistic beaming can shift the line center blue-ward. To shift the line center back to the location of the observed feature, the gravitational red-shift must be enhanced either by moving the emission region closer in, or by increasing the mass of the compact object.

We can take this knowledge and create a best fit to the observed feature of XTE J1810-197.  For a rotating shell, we get a best fit with a spherical shell having an inner and outer radius of 20 km and 25 km respectively, a compact object mass of $M=1.4M_\odot$  , a constant velocity of $v=0.2c$ and an inclination (of the rotation plane) of $32^\circ$.  The results are shown in Fig.\ref{fig:atomic_fit}.  Again this fit is not unique as, for example, a rotation in inclination is equivalent to decreasing the velocity.  The fit reproduces the width of the 13 keV feature quite well and also reproduces the small ``bump'' at $\sim$17 keV.  As mentioned earlier, this bump may or may not be real.  In either case, the Rb atomic line model does predict it at the given temperature. 

For the rotating disk case, we found the optimal parameters to be a disk with an inner and outer radius of 18 km and 30 km respectively and a thickness of 5 km.  The velocity was kept at $v=0.2c$, but the inclination was decreased to $25^\circ$.  The fit shown in Fig.\ref{fig:atomic_fit} also reproduces the observed width of the line quite well.  As with the spherical case, the ``bump'' at $\sim$17 keV is well replicated in this model. 

We know that when a rotating disk is seen edge on, the line is split into a red and blue component.  When relativistic beaming is significant, the red component is less intense than the blue.  We next consider that the observed feature $\sim$9 keV may in fact be this red component of the line split in the velocity field.  If this were true, the actual line center would not be at 13.07 keV, but $\sim$11 keV.  The feature at $\sim$9 keV would then be the red side of the split line, and the main feature at 13.07 keV would be the blue side of the feature.  In order to reduce the number of free parameters, we assume the disk is rotating with Keplerian velocity, that is:

\begin{equation}
v(r) = \sqrt{\frac{GM}{r}}
\label{eq:vkepler}
\end{equation}

With these criteria, we now have four free parameters for our model: inner and outer radius of the disk, mass of the compact object, and the inclination of the disk.  Our best fit model with these parameters is a disk with inner and outer radius of 12 km and 16 km respectively, inclined at $32^\circ$ around a compact object with $M=2M_\odot$.  The geometrical mesh representation of this model is shown in Fig.\ref{fig:atomic_mesh} and the resulting spectrum in the bottom right panel of Fig.\ref{fig:atomic_keplerian_disk_sequence} convolved with a Gaussian of width 1.5 keV to match the resolution of the observations.  Not only is the main feature at 13.07 keV fit very well, but the two smaller features at $\sim$ 9keV and $\sim$17keV are as well.  The feature at 9 keV is the red component of the split line, the feature at 13.07 keV is the blue component, and the feature at $\sim$17 keV is the contribution from the blue side of the weaker, but more abundant, Rb lines in this region.  To show the breakdown in processes contributing to the line shape, we provide Fig.\ref{fig:atomic_keplerian_disk_sequence}, a four panel figure showing the spectrum with increasing levels of physical accuracy.  We use 1000 frequency bands to show the fine structure of the spectrum.  The top left panel shows the original non-shifted spectrum composed of several Rb transitions.  The top right panel shows the effects of adding the gravitational field; the line is shifted to the red and slightly broadened.  The bottom left is the addition of a Keplerian velocity profile, which has the effect of splitting the line into a red and blue component.  Finally the bottom right panel adds the relativistic beaming effect which suppresses the red and enhances the blue component of the split line.  We should note that the observed features at $\sim$9 keV and $\sim$17 keV are not necessarily real as they have low significance.  Further, more sensitive observations of XTE J1810-197 are therefore needed to obtain better data which may resolve this issue.

These simulations have shown that atomic lines, specifically those of Rb, can explain the 13 keV feature seen in XTE J1810-197.  If the feature at $\sim$9kEv and $\sim$17keV are real, the model that best fits the data is a Keplerian disk with an inner and outer radius of 12 km and 16 km respectively in a gravitational field generated by a compact object of mass $M=2M_\odot$.  If these two smaller features are not real, the main feature at 13.07 keV can be well reproduced by a rotating disk, rotating sphere or a static extended atmosphere all in a weaker gravitational field.

\section{Discussion}

We have investigated how the 13 keV feature of XTE J1810-197 may be reproduced under different conditions and processes.  We now take this knowledge and apply it to the possible theories for AXPs.  For each emission process (cyclotron or atomic) we examine the feasibility for each AXP model.

\subsection{Cyclotron Emission}

Under the right conditions, both electron and proton cyclotron emission can reproduce the 13 keV feature in XTE J1810-197.  Within a dipole field, we found that emission from a thin region above the poles is sufficient.  For proton cyclotron radiation, the B field within this region must be $\sim 2\times10^{15}$ G and for electron cyclotron radiation, it must be $\sim 1\times10^{12}$ G.

In the Magnetar model, the surface is expected to have magnetic fields $B \sim 1\times10^{14} \to 1\times10^{15}$ G.  Cyclotron emission from the surface must therefore be proton in nature in order to account for the line center of the 13 keV feature.  This would support the theory that the burst originates from fractures in the neutron star crust.  If the emission is in fact electron cyclotron emission, it must come from a region considerably higher than the neutron star surface. Fig.\ref{fig:dipole_emission_region} (blue X’s) shows the possible electron cyclotron emission regions around a neutron star with a maximum field strength of $1\times10^{15}$ G on its surface ($\sim 12$ km).  We can see that the possible emission regions are between $\sim$90 km and $\sim$120 km.  This would support the idea that the burst was caused by a reconnection event in the magnetosphere.  It is difficult to determine whether the feature is caused by proton or electron cyclotron emission since the width of the line is dominated by the variability in the magnetic field, and not by any differences between the two particles. The spin down rate of XTE J1810-197 implies a magnetic field of $B \sim 3\times10^{14}$ G where the proton cyclotron interpretation requires a much larger field of $B \sim 2\times10^{15}$ G.  This discrepancy may therefore favour the electron cyclotron model.

The magnitude of the magnetic field in the FBD model is expected to be $< 1\times10^{13}$ G .  This is much too low for proton cyclotron emission at the required energies and therefore only electron cyclotron radiation from the surface of the neutron star is feasible in this model.  Electron cyclotron radiation from the disk is ruled out since the field at the disk location ($\sim$10,000 km) is much too weak.

In the QN model, only electron cyclotron emission is plausible.  The QN remnant is a quark star  in a color-superconducting state \citep{vogt_2004, ouyed_2005} and is not expected to have a crust \citep{alford_1999}.  So although the field strengths at the surface ($B \sim 1\times 10 ^{15}$ G) are enough to provide the necessary proton cyclotron energies, no emission from this region will take place.  Electron cyclotron emission from the Keplerian ring, however, may be possible.  Fig.\ref{fig:dipole_emission_region} (red X’s) shows the possible electron cyclotron emission regions around a quark star with a maximum magnetic field of $B \sim 1\times 10 ^{15} G$ on its surface (radius of 8 km).  If electron cyclotron emission around a quark star is responsible for the 13 keV feature, it must be emitted from a region between $\sim$60 km and $\sim$83 km.  The Keplerian ring in the QN model is expected to be located between $\sim$20 km and $\sim$100 km, which makes electron cyclotron radiation in this scenario a real possibility.

Cyclotron emission in each model therefore appears to be a viable cause of the 13 keV feature seen in XTE J1810-197.  If XTE J1810-197 was the only available spectrum, this would be an appropriate conclusion.  However, emission lines have been observed in two other AXPs, 4U 0142+61 and 1E 1048.1, which also show a significant feature at $\sim$13 keV (Fig.\ref{fig:14kev_lines}).  As we have seen from our simulations, the location of the cyclotron line depends on the magnetic field strength within the emitting region and the inclination of that region with respect to the observer.  For the feature to appear in each AXP spectrum, the emission region in each must therefore have the same magnetic field and orientation. Or, which is even more unlikely, a suitable combination of the two that coincidentally results in the same line center.

\subsection{Atomic Line Emission}

We have shown that under various conditions (i.e. rotating disk, rotating shell, extended atmosphere) atomic line emission from Rubidium can adequately reproduce the 13 keV feature of XTE J1810-197.

\subsubsection*{Rubidium}

Rubidium isotopes $^{85}$Rb and $^{87}$Rb are produced through the slow (s-) and rapid (r-) nuclear processes.  The s-process occurs in low mass asymptotic giant branch stars (AGB) \citep{garcia_2006} and the r-process likely during a supernova or quark nova explosion.  In a supernova, much of the Rb is distributed into the interstellar medium, whereas in a quark nova the Rb will be retained in the Keplerian ring (depending on the quark nova parameters).

In the case of a magnetar, most Rb is likely removed from the system, although some may remain within the crust.  If the ``starquake'' produces fractures in the neutron star crust, this Rb may be able to make its way to the surface and emit.  The gravitational field produced by the neutron star must be such that the Rb line is shifted from its emitted energy of 14.299 keV to the observed energy at 13.07 keV.  The mass of the neutron star required for the Rb emission from a typical radius of 10 km is $\sim 0.6M_\odot$.

In the FBD model, the Rb expelled during the SN explosion may fall back and make up some of the disk.  It is possible, therefore, for Rb to be present in the spectrum.   For the accretion scenario, the shift of the line requires a neutron star with a mass of $\sim 0.6M_\odot$ for an emitting radius (surface) of 10 km.

The QN is an explosion of the neutron star crust resulting in an abundance of iron and neutrons.  The r-process is therefore very efficient during a QN explosion.  We have performed calculations using the code r-Java \citep{char_2011} to calculate the abundance of Rb under the QN scenario and find that Rb is readily produced in the QN ejecta.  Since the QN ejecta form the Keplerian ring, we can expect a significant abundance of Rb available to emit.

\subsubsection*{Broadening} 

In order to explain the breadth of the 13 keV feature with Rb atomic lines, the atoms must be either moving a significant fraction of the speed of light, or extended around a compact massive object.  For the Magnetar model, the Rb from presumably the surface would have to be travelling $\sim$0.2c.  Given that the rotation period of XTE J1810-197 is $\sim$5.5 s, the radius of the neutron star would have to be $\sim$5200 km to attain these speeds on its surface.

This same argument holds for the accretion onto the neutron star in the FBD model. The emission, however, may come from the disk itself.  If the disk were Keplerian, we can see from equation \ref{eq:vkepler} that for a velocity of 0.2c and an inner radius of $\sim$10,000 km the mass of the compact object would have to be at least $196 M_\odot$, again unreasonable.

The Rb emission in the QN model is expected from a Keplerian ring with an inner and outer radius between $\sim$20 km and 100 km respectively.  We have already modelled this situation, and found that a Keplerian disk with an inner and outer radius of 12 km and 16 km respectively around a compact object with a mass $M=2M_\odot$ fits the observations very well (Fig.\ref{fig:atomic_keplerian_disk_sequence}).

Gravitational broadening of the Rb lines requires an extended atmosphere.  For a compact object with a mass of  $1.4M_\odot$, the inner and outer edge of the atmosphere must be $\sim 13$ and $> \sim 300$ km respectively.  The emission, if the breadth is due to gravitational broadening, can therefore not come from the surface of the neutron star.  This precludes the FBD theory, assuming the emission is from accretion onto the star.  Emission from the disk (inner radius $\sim$ 10,000 km) is also unlikely because the mass of the compact object required to broaden the line would be $\sim 714M_\odot$.  In the Magnetar model, emission of Rb from the surface of the star is discarded using the same argument.  However, emission may come from the atmosphere well above ($\sim$100 km) the magnetar surface.  Ibrahim et al (2001) suggest that radiation pressure may lift a thin layer of material away from the magnetar surface.  Whether or not enough Rb could be driven off the surface, or if it could even reach an altitude of $\sim$100 km or more is uncertain.  The observation of the 13 keV feature in the tail of the burst does, however, seem to support the idea. Because of this we cannot fully rule out gravitational broadening in the Magnetar model. The QN model predicts an inner disk radius close to the required value of $\sim$13 km. \citet{ouyed_2007b} provides an expression for the outer radius of the Keplerian ring as a function of equilibrium temperature and age.  For an age of $\frac{P}{3\dot{P}} \approx 8720$ yrs and an equilibrium ring temperature of 0.5 keV, we calculate an outer radius of $\sim$306 km which agrees with the required radius of $>\sim$300 km.  Gravitational broadening in the QN model can therefore explain the full width of the 13 keV feature in XTE J1810-197, although it is more likely that both Doppler and Gravitational broadening contribute.

\section{Conclusion}

\subsection{Summary}

We have performed an analysis of different geometrical and physical conditions required to explain both the location and shape of the 13 keV feature seen in XTE J1810-197.  Both cyclotron (positron and electron) and Rb atomic line emission are possible explanations for this particular feature.  

If the feature is proton cyclotron emission, the only model that can satisfactorily explain it is the Magnetar model.  In this case the emission must be coming from the surface of the neutron star.  Electron cyclotron emission can be explained in the Magnetar, FBD, and QN models.  In the Magnetar model, the electron emission must come from well above the neutron star, perhaps in the magnetosphere. Due to the low surface magnetic fields expected in the FBD model, electron cyclotron radiation would have to come from the stellar surface.  In the QN model, the electron emission would come from the Keplerian ring where the magnetic fields are lower.

Atomic line emission from Rb is also possible in each model.  Rb is the most likely species responsible for the 13 keV feature, and is expected to exist, at least minimally, in each theory.  The Magnetar and FBD models predict small traces of Rb whereas the QN theory predicts large quantities.  The breadth of the Rb line requires velocities approaching the speed of light or extended atmospheres, both unexpected in the Magnetar model.  The fall-back disk of the FBD model is too large for either of these processes to adequately explain the shape of the 13 keV feature.  This does not, however, preclude the presence of a fall-back disk around this, or any other AXP source.  The FBD model requires an additional mechanism to explain the bursting, which may also provide a site for the emission of the 13 keV feature.  The QN model, on the other hand, predicts a close, rapidly rotating Keplerian ring which is more than adequate to explain the breadth in terms of Doppler broadening alone.  Gravitational broadening may also contribute to the breadth of the line in this situation.  The only theory that can satisfactorily account for all properties of the 13 keV feature under the atomic line hypothesis is that of the quark star in the context of the QN explosion.  

If XTE J1810-197 were the only source with the 13 keV feature, cyclotron emission would be a very likely explanation.  However, in every AXP with an emission spectrum a feature with the same line center has been observed.  This would require all three sources to have emission regions with the same magnetic field and orientation with respect to earth.  Although possible, this scenario is not very likely.  For this reason we do not believe that the 13 keV feature in XTE J1810-197 is cyclotron in nature, and is more likely to be an atomic transition of Rb.  A summary of our finding is presented in Table \ref{table:compare}.

\subsection{Predictions}

We have found that the QN model provides the best explanation for the location and breadth of the 13 keV feature in XTE J1810-197.  We can therefore make several predictions based on this model:

\begin{itemize}
\item The Keplerian ring in the QN model is unique to AXPs and will split the atomic lines into a red and blue component.  SGRs, under this model, are explained by a co-rotating shell at much lower velocities.  We therefore do not expect to see any line splitting in SGRs.

\item We expect to see a feature at $\sim$17 keV representing the less intense, but more abundant Rb lines.  A strong detection of this feature would support this model.  A null detection, however, may not rule it out since the strength of these lines depends on the excitation conditions (i.e. temperature and density).

\item The relative strength of the $\sim$17 keV feature and the main $\sim$13 keV feature is dependent on the relative abundance of RbXXXVI and RbXXXVII (the former has no lines around 17 keV).  A strong detection of the $\sim$17 keV feature will therefore provide constraints on the electron density and temperature of the emitting region.

\item If atomic lines are responsible for the 13 keV feature, we should see evidence of other lines during quiescence, in the form of absorption features.

\item During and shortly after a burst, the atmosphere around the Keplerian ring is expected to be abundant in both ionized heavy-elements and electrons.  Since the magnetic field penetrates the ring atmosphere, we expect to see electron cyclotron lines along-side atomic lines in the spectra of future AXP observations.

\item The width of the 13 keV feature in XTE J1810-197 is not well constrained \citep{woods_2005}.  Future, high resolution, observations will provide better data  allowing us to more accurately determine the required velocities.  This will put strong constraints on AXP models, and provide invaluable information regarding the environment around these enigmatic objects.

\end{itemize}

\acknowledgments

This work is funded by the Natural Sciences and
Engineering Research Council of Canada.  N.K. would like to acknowledge support from the Killam Trusts.




\clearpage


\begin{onecolumn}

\begin{figure}
\resizebox{\hsize}{!}{\includegraphics{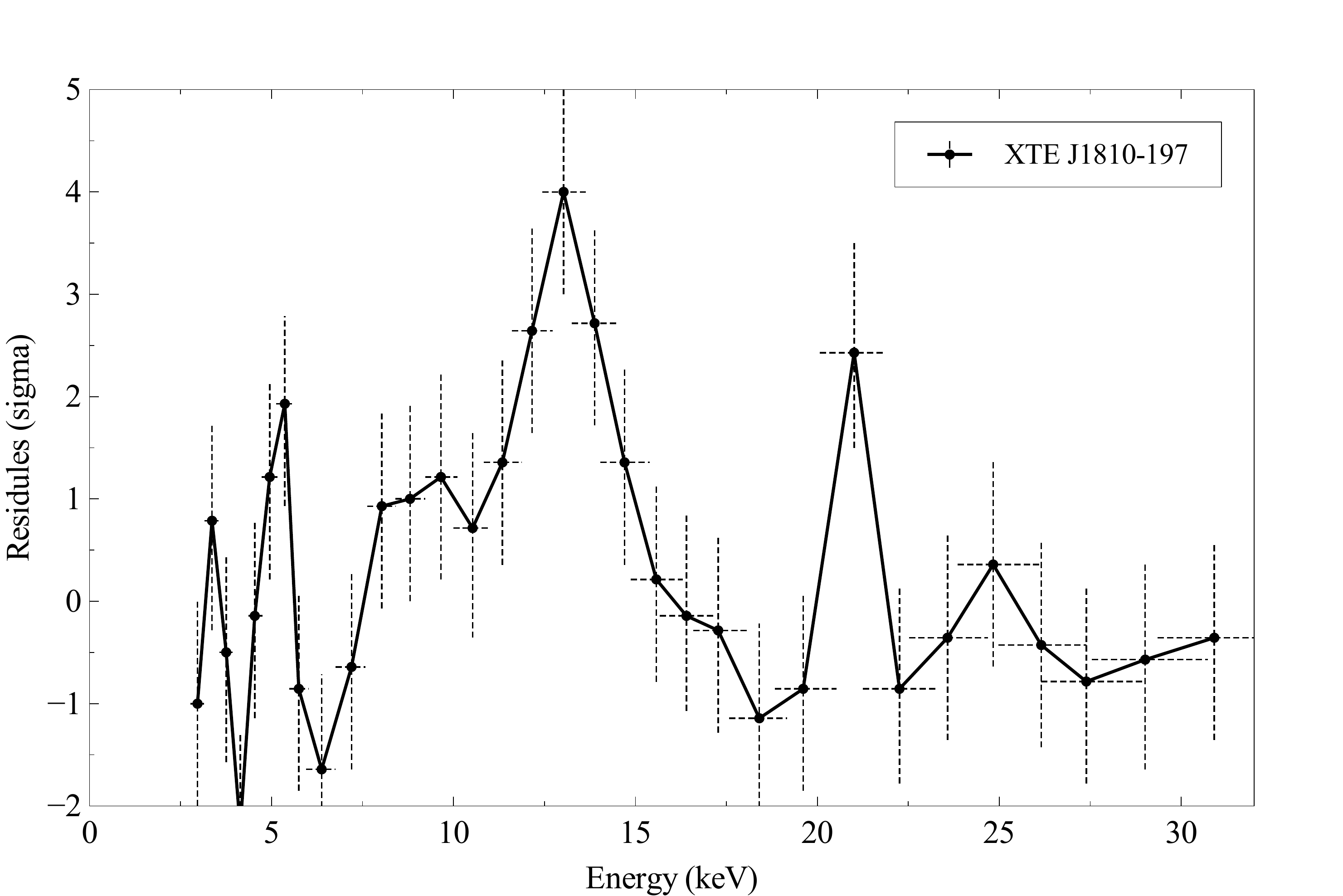}}
\caption{Observed spectrum of XTE-J1810-197 using data from \citet{gavriil_2008}.  The y-axis represents the residuals after the black-body model fit is subtracted from the original data (see \citet{gavriil_2008}).
}
\label{fig:observed}
\end{figure}

\begin{figure}
\resizebox{\hsize}{!}{\includegraphics{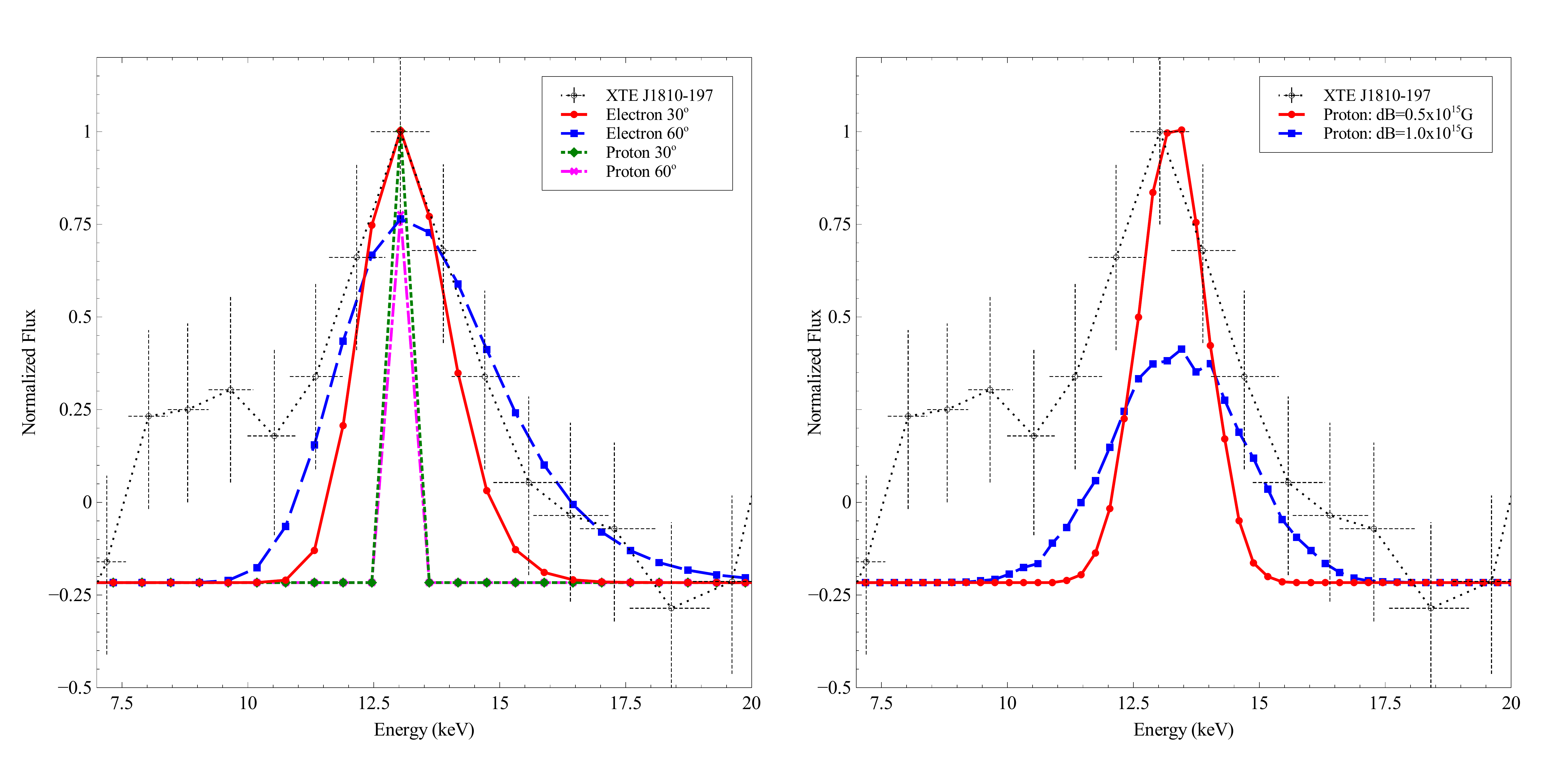}}
\caption{Left: Cyclotron emission lines in a constant magnetic field of $1\times10^{12}$ G and $2.1\times10^{15}$ G for the electron and proton cases respectively.  Angles at $30^\circ$ and $60^\circ$ between the magnetic field and observer are shown, as well as the observed spectrum of XTE J1810-197.  Right: Proton cyclotron emission lines in a variable magnetic field;  $B = 2.1\times10^{15}$ G $\pm dB$.  Two values of $dB$ are shown, as well as the observed spectrum of XTE J1810-197.  It is clear that the larger variation in magnetic field produces the broader line.
}
\label{fig:ep_cylo_Bconstant}
\end{figure}

\begin{figure}
\resizebox{\hsize}{!}{\includegraphics{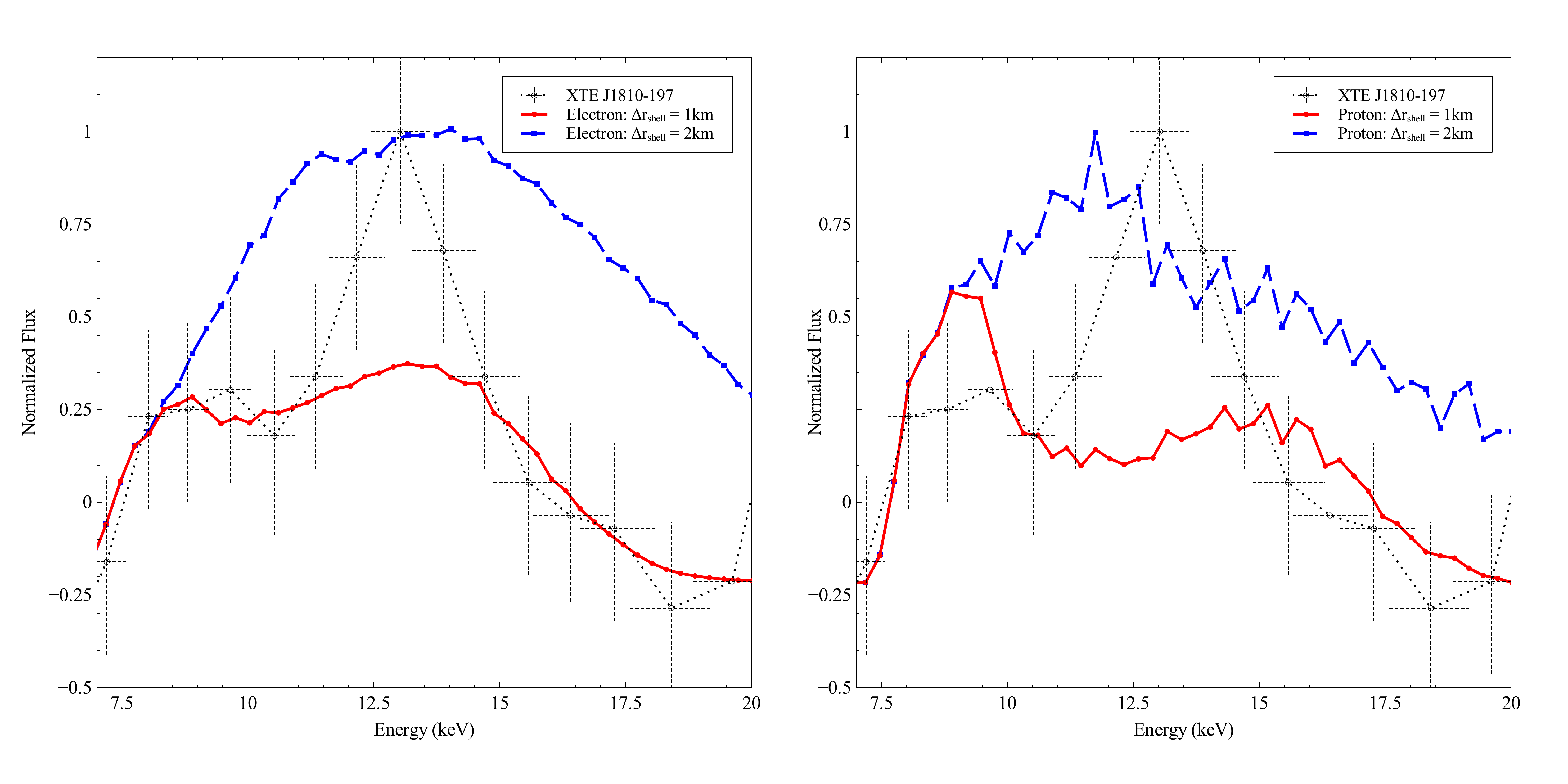}}
\caption{Cyclotron emission from a spherical shell within a dipole magnetic field.  Two different shell widths are plotted, 1km and 2km.  The outer radius of the shell in both cases is 12 km.  The observed spectrum of XTE J1810-197 is plotted in the background for reference.  Left: Electron cyclotron model.  Right: Proton cyclotron model.
}
\label{fig:ep_cylo_shell}
\end{figure}

\begin{figure}
\resizebox{\hsize}{!}{\includegraphics{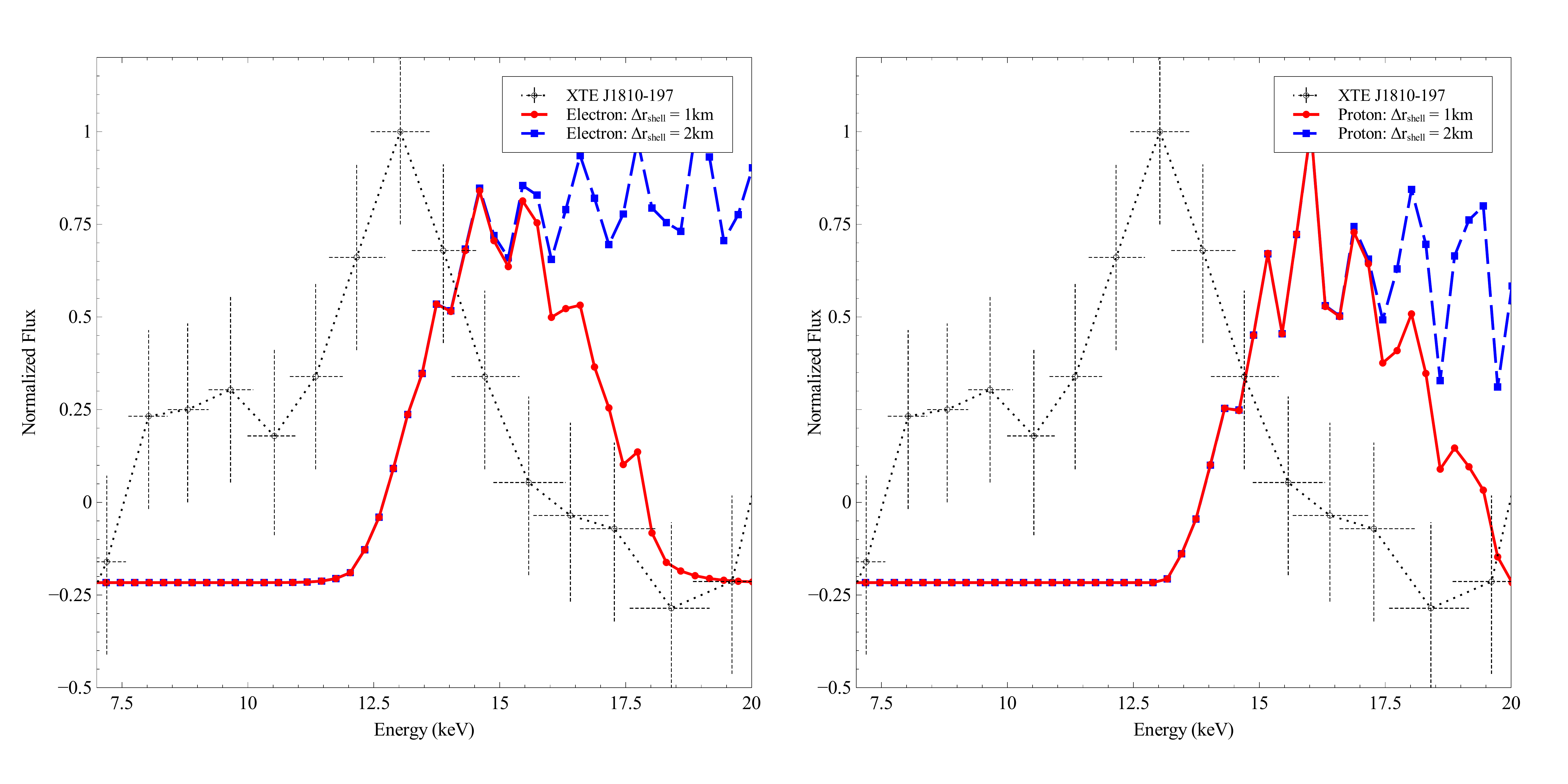}}
\caption{Electron cyclotron emission from the polar regions between $\phi=0^\circ$ and $\phi=20^\circ$ of a spherical shell within a dipole magnetic field.  Two different shell widths are plotted, 1 km and 2 km.  The outer radius of the shell in both cases is 12 km.  The observed spectrum of XTE J1810-197 is plotted in the background for reference.  Left: Electron cyclotron model.  Right: Proton cyclotron model.
}
\label{fig:ep_cylo_pole}
\end{figure}

\begin{figure}
\resizebox{\hsize}{!}{\includegraphics{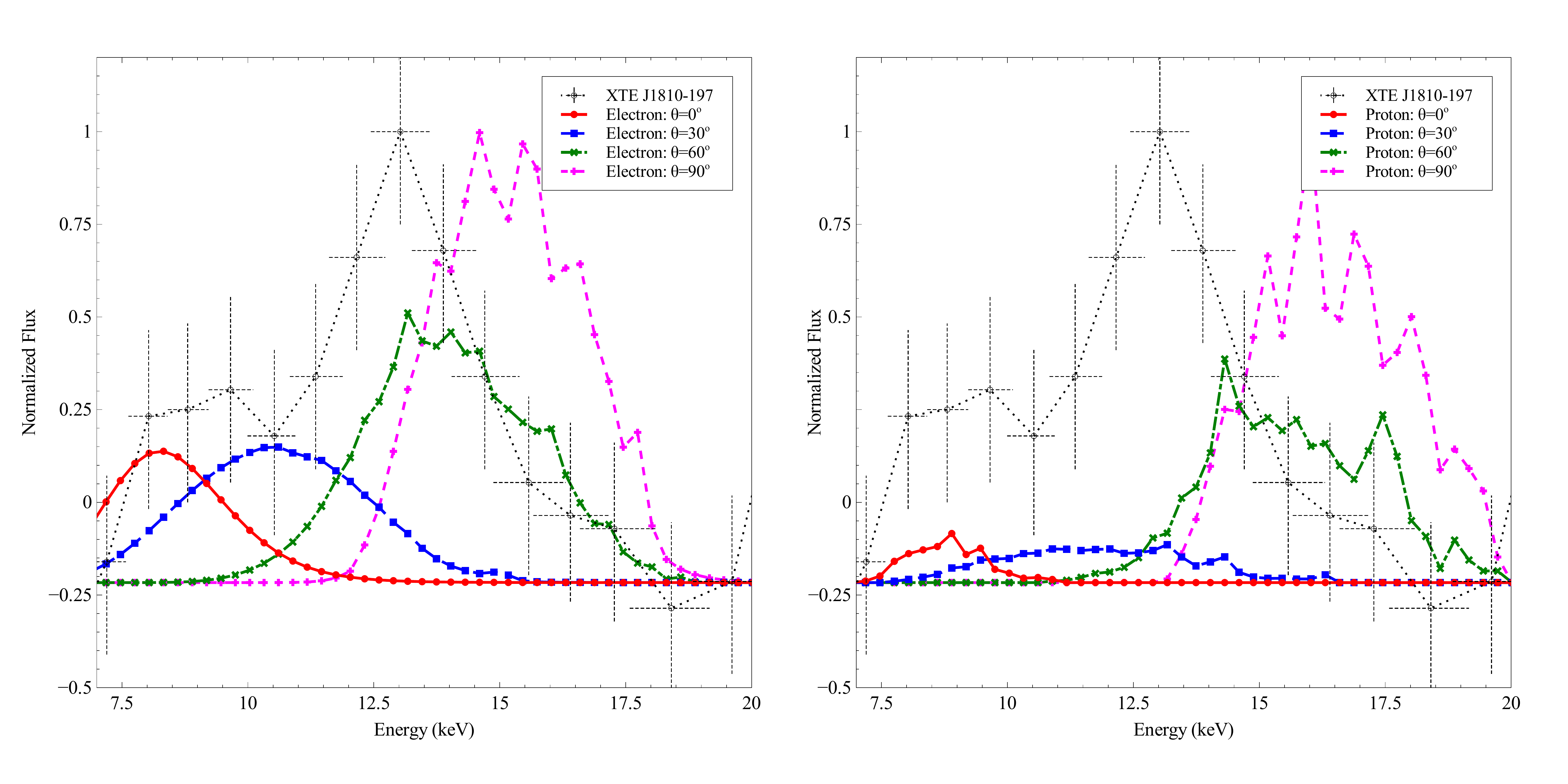}}
\caption{Electron (left panel) and proton (right panel) cyclotron emission from the polar regions between $\phi=0^\circ$ and $\phi=20^\circ$ of a spherical shell within a dipole magnetic field.  Plotted are the results of simulations of four different angles between the magnetic moment of the dipole and the observer.  In all cases the thin atmosphere of 1 km is used. The observed spectrum of XTE J1810-197 is plotted for reference. 
}
\label{fig:ep_cylo_poleangles}
\end{figure}

\begin{figure}
\resizebox{\hsize}{!}{\includegraphics{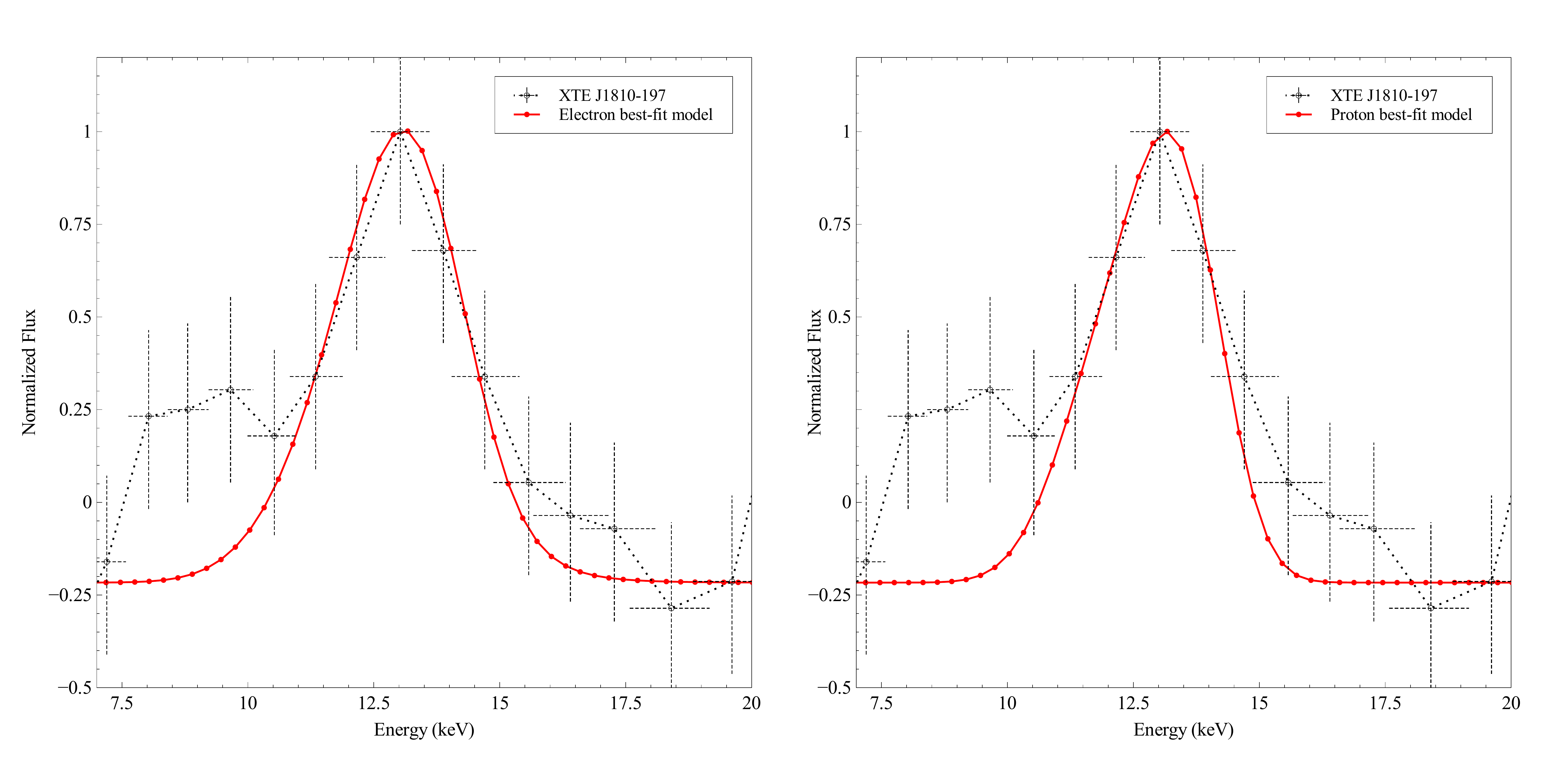}}
\caption{Best fit model to the observed 13 keV feature of XTE J1810-197 for cyclotron emission.  The emission is from a thin spherical shell with an inner and outer radius of 11.9 km and 12 km respectively, centred around the pole ($\phi=0^\circ \to 20^\circ$) of a magnetic dipole field.  Left: Electron cyclotron model with a dipole magnetic moment of $\mu_m=1\times10^{27}$ A m$^2$ at an inclination of $66^\circ$ to the observer.  Right:  Proton cyclotron model with a dipole magnetic moment of $\mu_m=2\times10^{30}$ A m$^2$ at an inclination of $56^\circ$ to the observer.  Each spectrum is convolved with a Gaussian of width 1.5 keV to simulate the energy resolution of the observations.
}
\label{fig:ep_cylo_bestfit}
\end{figure}

\begin{figure}
\resizebox{\hsize}{!}{\includegraphics{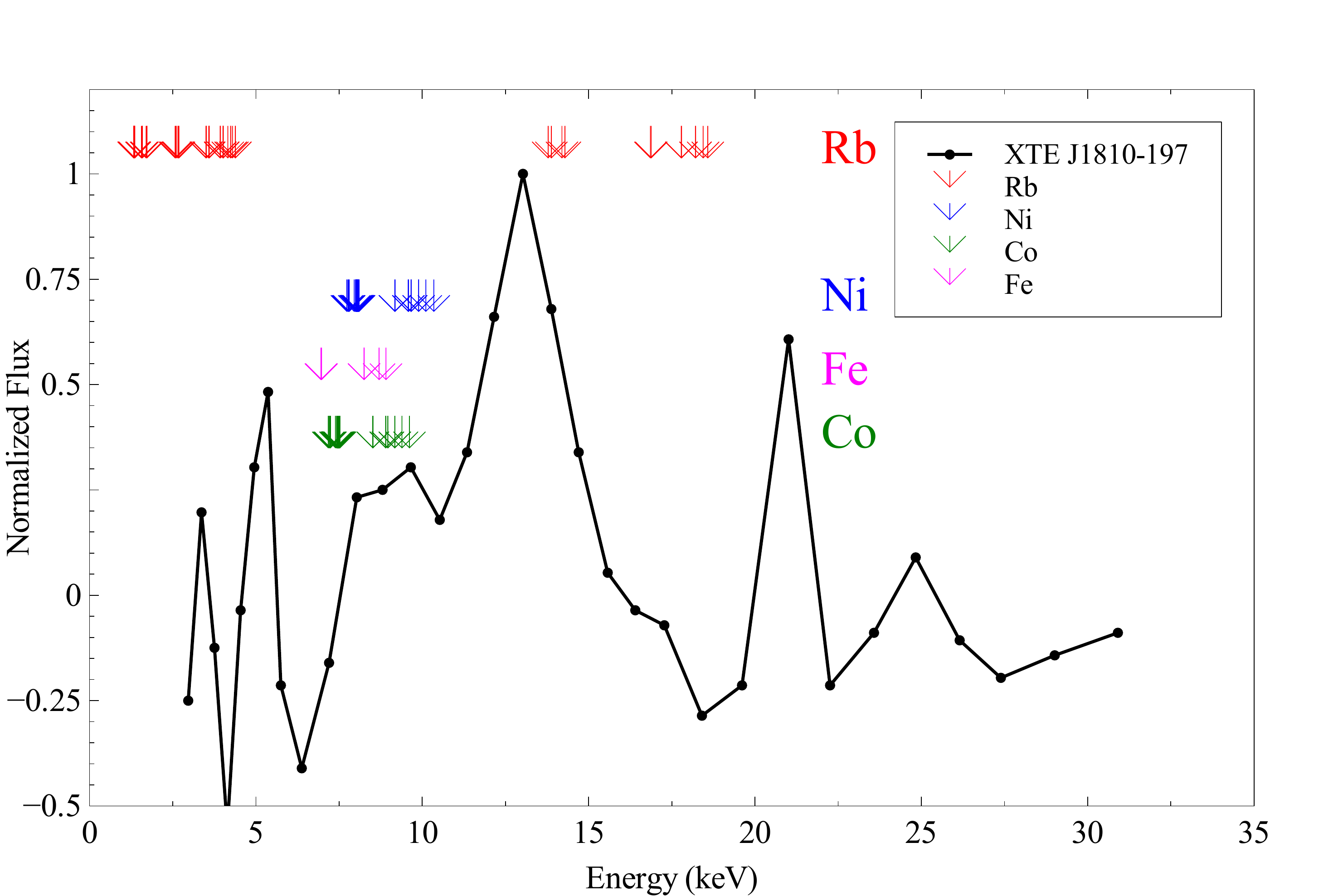}}
\caption{Atomic transitions near the 13 keV feature of XTE J1810-197}
\label{fig:atomic_identification}
\end{figure}

\begin{figure}
\resizebox{\hsize}{!}{\includegraphics{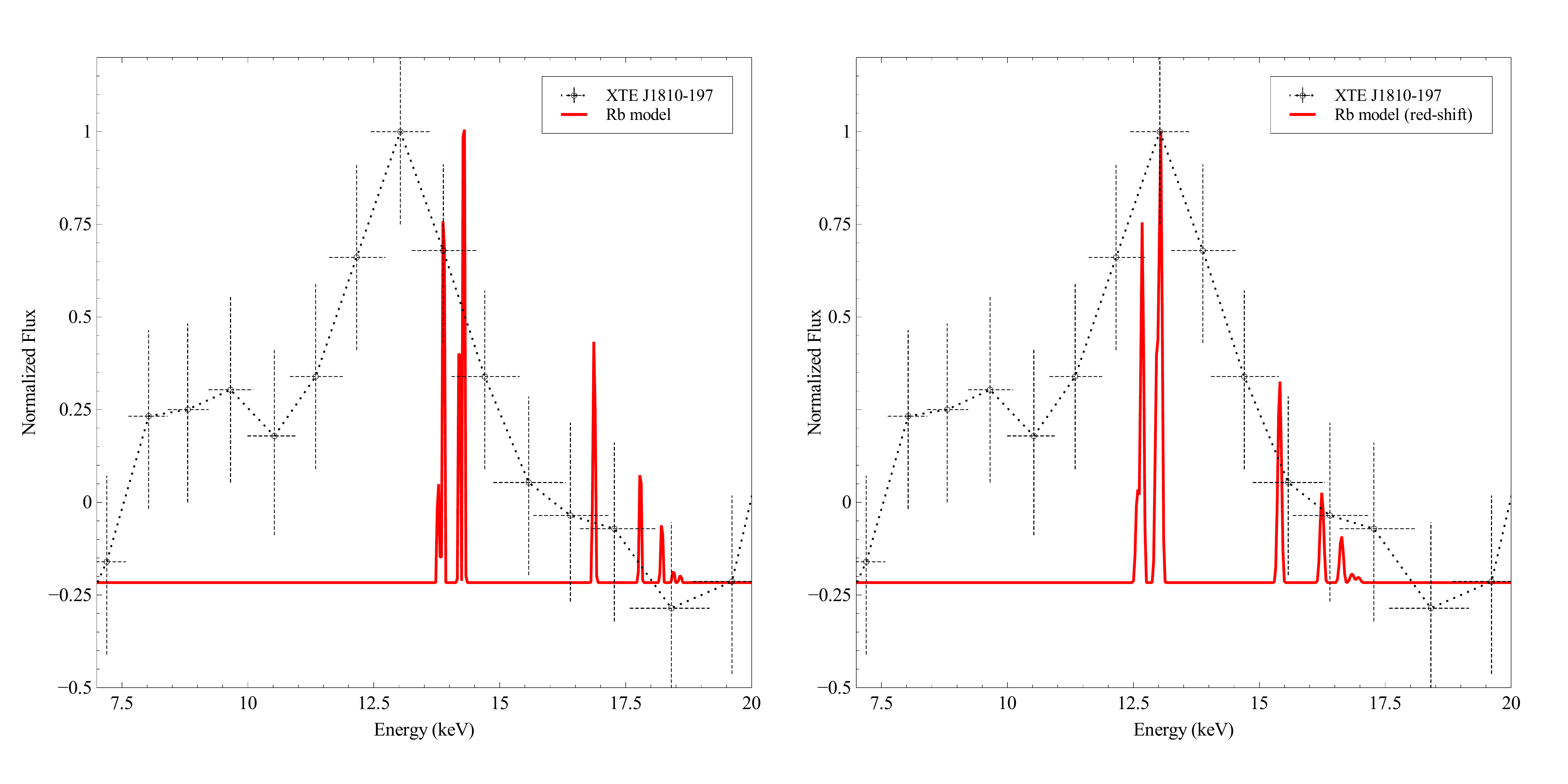}}
\caption{Model spectrum of Rb transitions at $T=1.5\times10^8$ K.  Rb XXXVII and Rb XXXVI transitions are shown.  The abundance of RbXXXVI is 10\% that of RbXXXVII. 1000 frequency bands were used to show the fine structure. Left: Original, unshifted spectrum.  Right: Gravitationally shifted lines from a $1.4 M_\odot$ compact object.  The emission is from a spherical shell with an inner and outer radius of 24.5 and 25.5 km respectively.
}
\label{fig:atomic_rb_lines}
\end{figure}

\begin{figure}
\resizebox{\hsize}{!}{\includegraphics{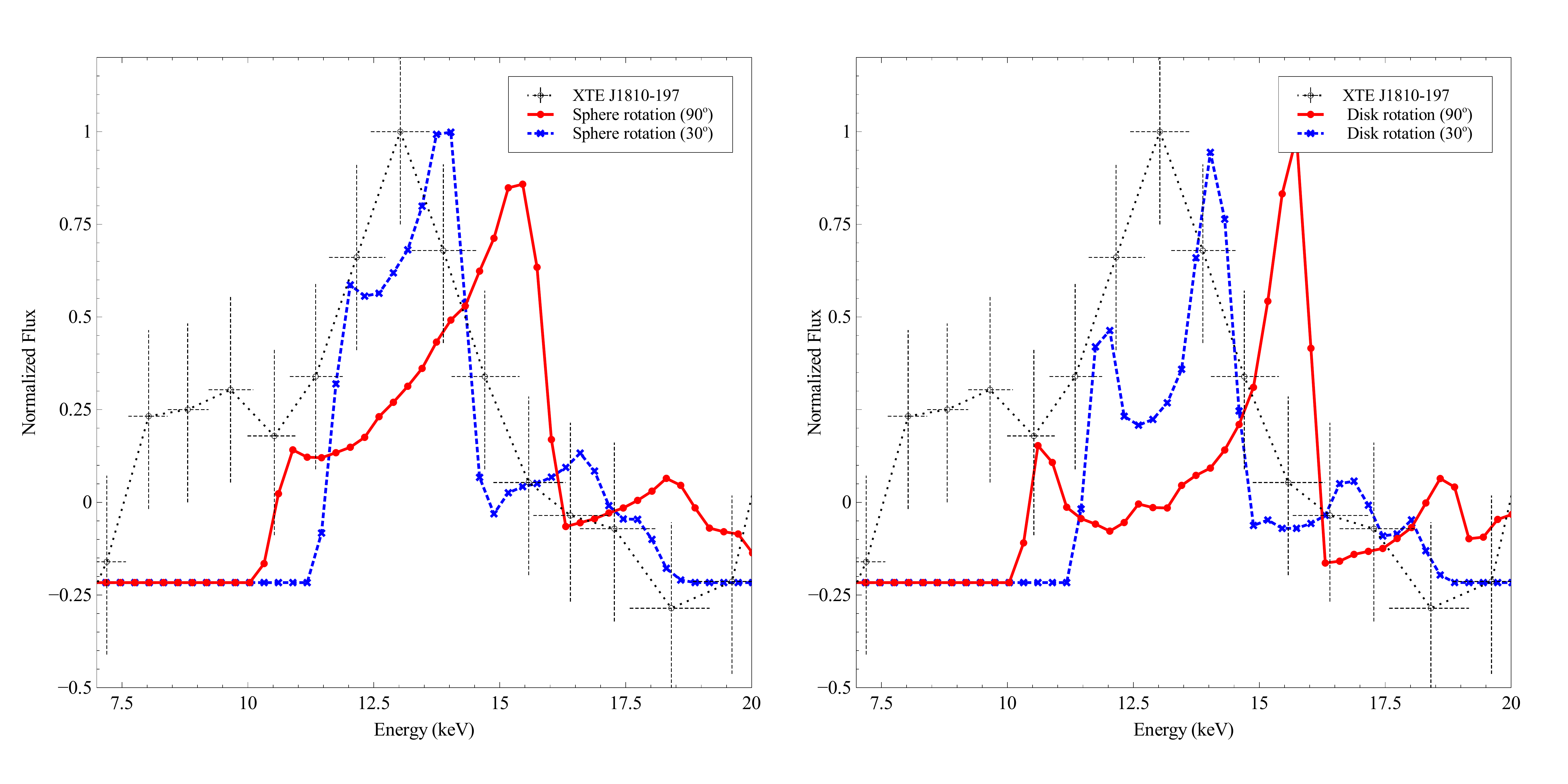}}
\caption{Constant velocity, Rb atomic line model.  Left: A spherical shell with an inner and outer radius of 24.5 and 25.5 km respectively.  Right:  A disk with an inner and outer radius of 24.5 and 25.5 km respectively.  The rotational velocity is $v=0.2c$ in both cases.  Two different inclination of the rotational plane are plotted for each model.
}
\label{fig:atomic_rotation}
\end{figure}

\begin{figure}
\resizebox{\hsize}{!}{\includegraphics{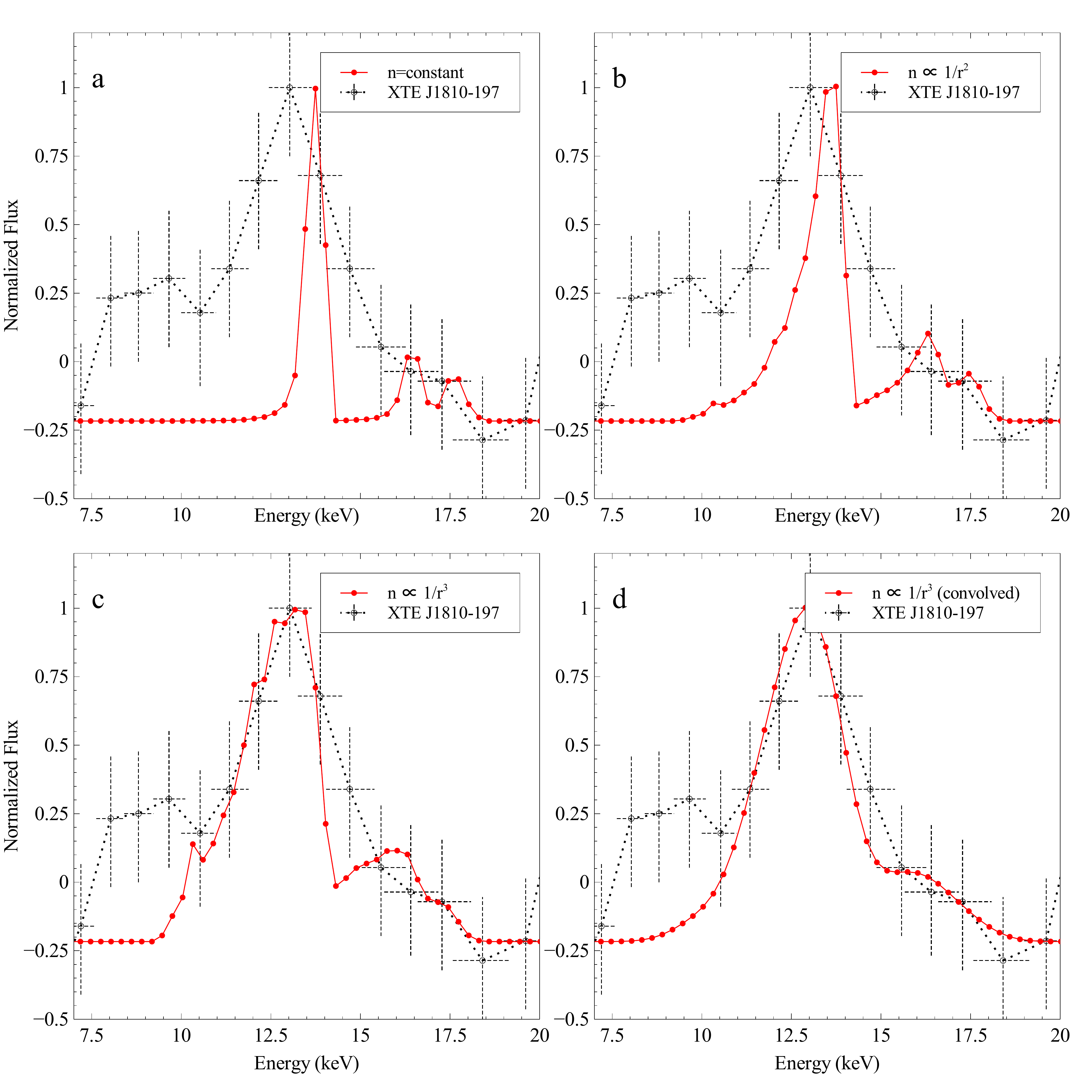}}
\caption{Gravitational broadening from a disk in a gravitational field.  The inner and outer radius of the disk is 10 km and 300 km respectively, and the mass of the compact object is $2M_\odot$.  a)  Broadening of the line with a constant density profile throughout the disk.  b) Broadening of the line with a $1/r^2$ density profile.  c) Broadening of the line with a $1/r^3$ density profile.  d)  Broadening of the line with a $1/r^3$ profile but convolved with a Gaussian of width 1.5 keV to simulate the energy resolution of the observations.
}
\label{fig:atomic_gravbroad}
\end{figure}

\begin{figure}
\resizebox{\hsize}{!}{\includegraphics{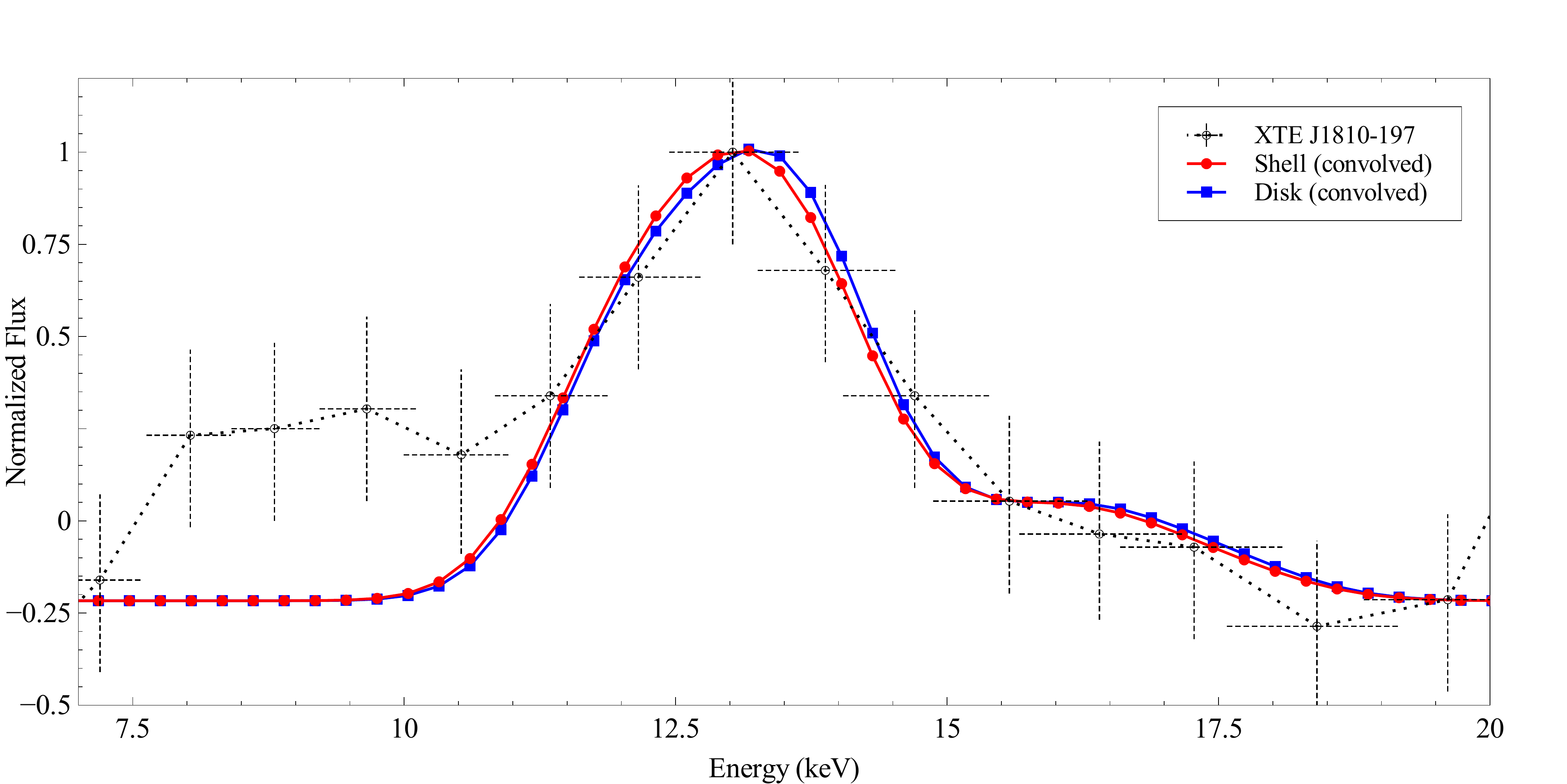}}
\caption{Best fit models under the atomic line model for a rotating spherical shell (red) and rotating disk.  In each model, a constant rotational velocity of $v=0.2c$ is used.  The data points were convolved with a Gaussian of width 1.5 keV to match observations.
}
\label{fig:atomic_fit}
\end{figure}

\begin{figure}
\resizebox{\hsize}{!}{\includegraphics{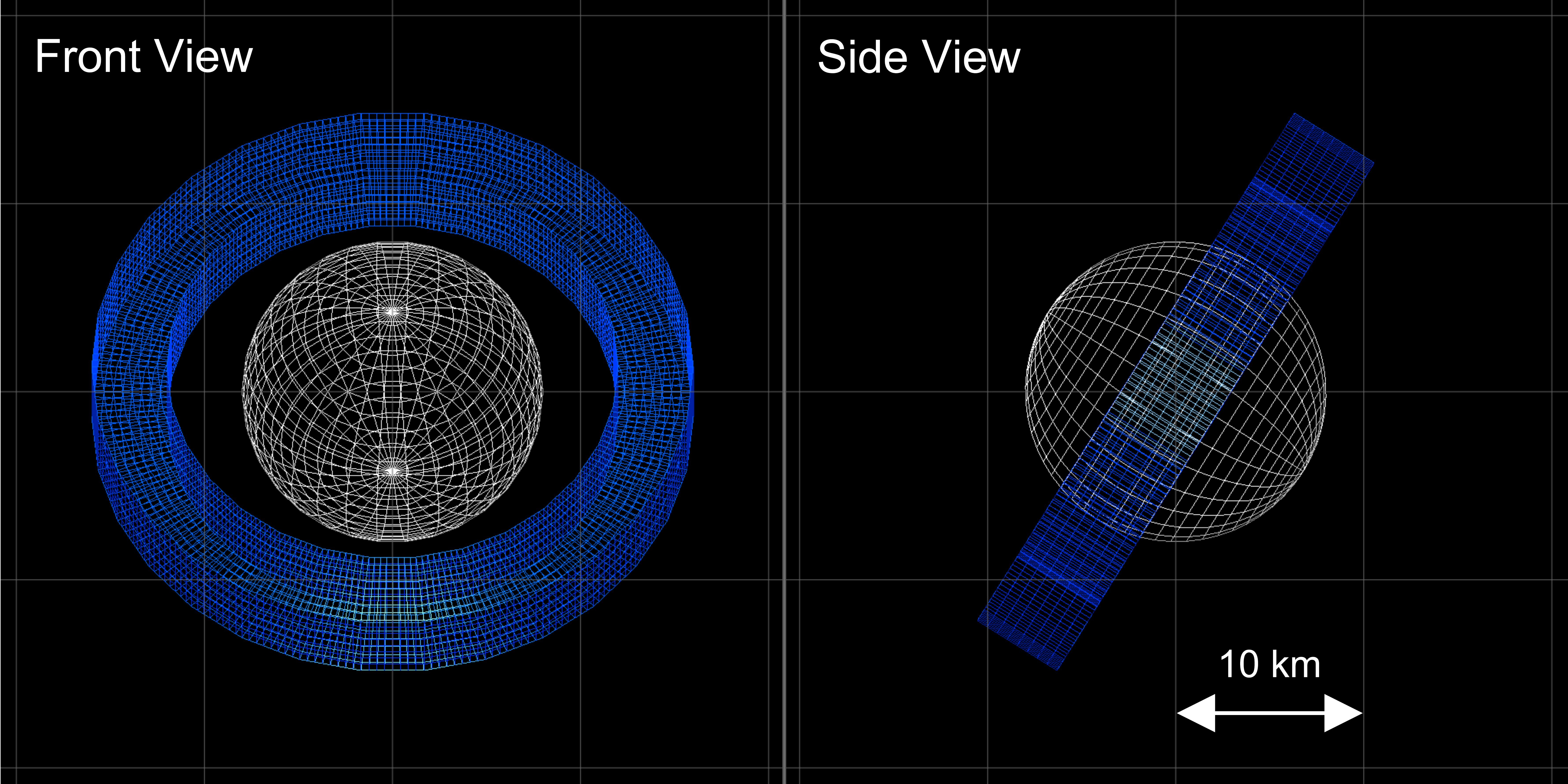}}
\caption{Geometrical representation of the Keplerian disk model.  The mass of the central object is $2M_\odot$, the inner and outer radius of the disk is 12 km and 16 km respectively.    The system is inclined at $32^\circ$.  Two different viewpoints are shown, the left panel from the front (observer) and the right panel from the side.  The compact object is shown with a radius of 8 km, and the grid is spaced in 10 km intervals.
}
\label{fig:atomic_mesh}
\end{figure}

\begin{figure}
\resizebox{\hsize}{!}{\includegraphics{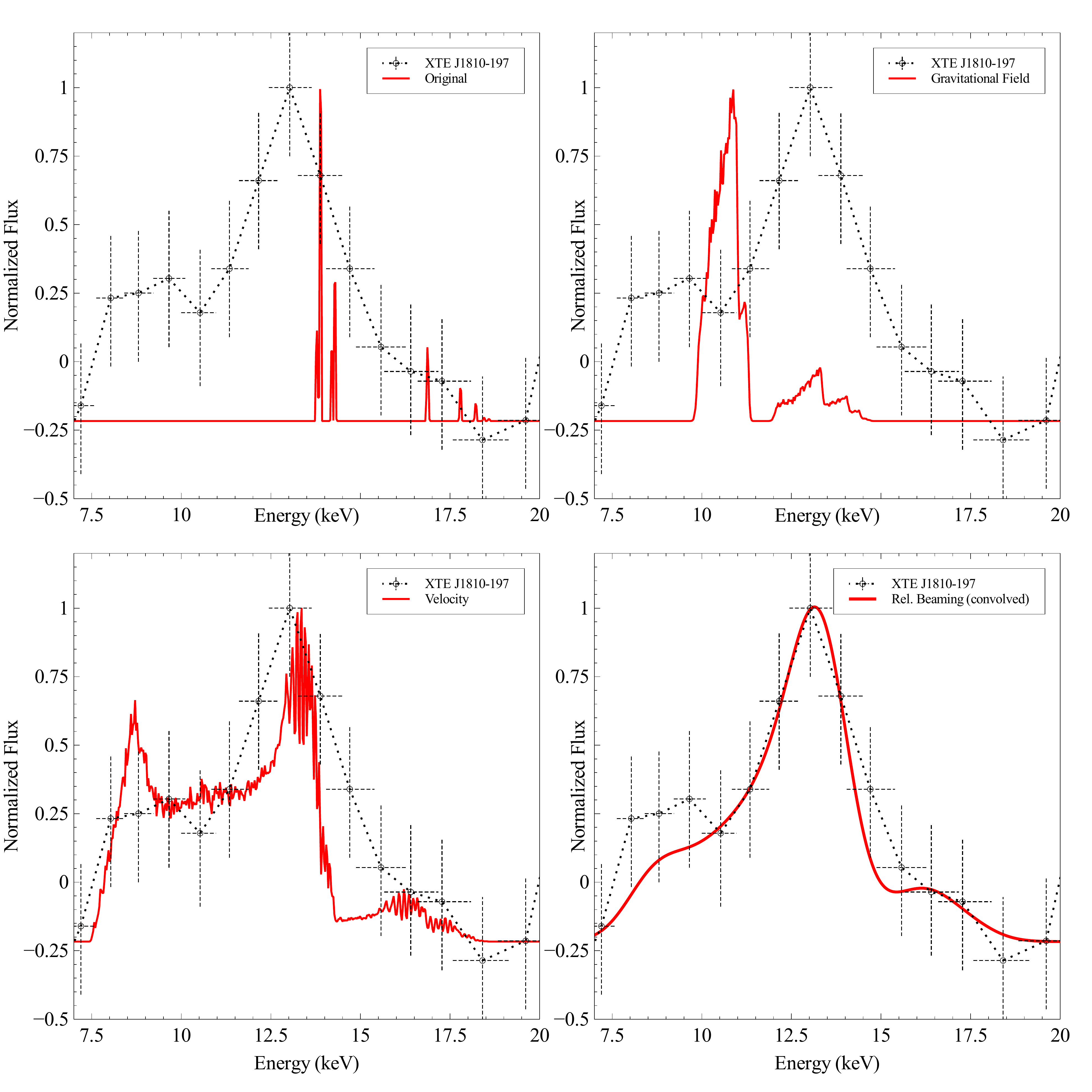}}
\caption{Sequence of panels showing the evolution of the line shape of the best fit atomic line as more physical processes are added. 1000 frequency bands were used to show the fine structure of the spectrum. The model consists of a Keplerian disk with an inner and outer radius of 12 and 16 km respectively, an inclination of $32^\circ$ and a compact object of $2M_\odot$.  Top Left:  Original, unmodified line shape contributed to by Rb transitions.  Top Right:  Gravitational field is added, causing the line to shift and broaden.  Bottom Left:  Keplerian velocity is added to the disk causing the line to split into a red and blue component of roughly equal intensity.  Bottom right:  Application of relativistic beaming causes the red side of the split line to be suppressed while the blue is enhanced.  The final panel represents our best fit Keplerian disk model, convolved with a Gaussian of width 1.5 keV to simulate the energy resolution of the observations.
}
\label{fig:atomic_keplerian_disk_sequence}
\end{figure}

\begin{figure}
\resizebox{\hsize}{!}{\includegraphics{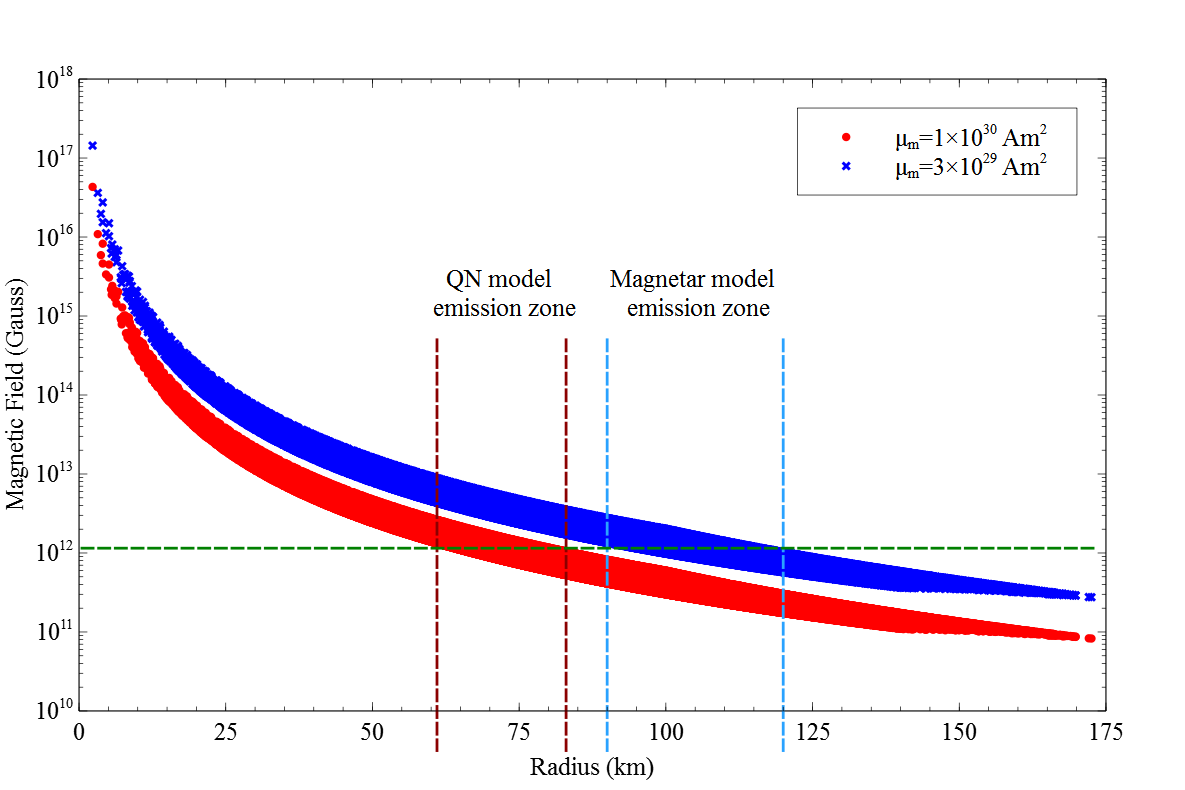}}
\caption{Plot of magnetic field strength with distance from an ideal magnetic dipole.  Two different field strengths are shown.  The blue xs represent the Magnetar model where the maximum magnitude at 12 km (assumed neutron star surface) is $1\times10^{15}$ G.  The red dots represents a quark star with a maximum magnitude of $1\times10^{15}$ G at its surface of 8 km.  The green dotted line shows the required field for electron cyclotron emission to account for the position of the 13 keV feature.  We can see that the emission region for the QN model is closer in than for the Magnetar model.
}
\label{fig:dipole_emission_region}
\end{figure}

\begin{figure}
\resizebox{\hsize}{!}{\includegraphics{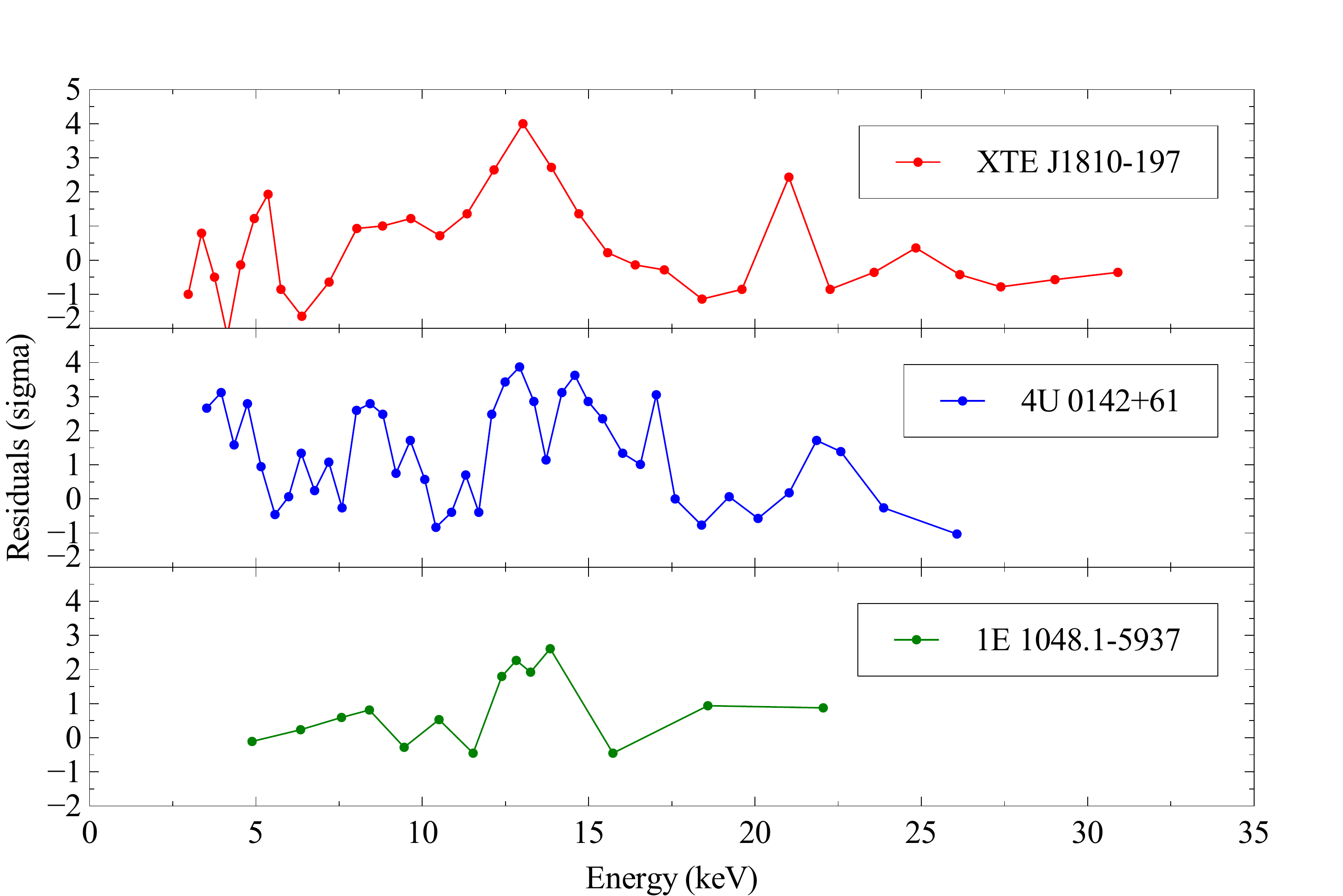}}
\caption{Observed emission line spectrum from three different AXP sources: XTE J1810-197, 4U 0142+61 and 1E 1048.1-5937.  All show a strong, significant feature at $\sim$ 13 keV.  Data from \citet{gavriil_2008}.
}
\label{fig:14kev_lines}
\end{figure}

\clearpage







\clearpage

\begin{deluxetable}{lccccc}
\tablecaption{Atomic transitions between $\sim$9 keV and $\sim$18.5 keV from the NIST database.\label{table:transitions}}
\tablewidth{0pt}
\tablehead{
\colhead{Species} & \colhead{Energy (keV)} & \colhead{$A_{ki}$} & \colhead{Config} & \colhead{Terms} &
\colhead{$g_i - g_k$}
}
\startdata

Rb XXXVII &	18.596&	2.24E+13 & 1s - 7p & 2S  - 2P* & 2 - 2  \\ 
Rb XXXVII &	18.457&	3.58E+13 & 1s - 6p & 2S  - 2P* & 2 - 2  \\ 
Rb XXXVII &	18.231&	6.40E+13 & 1s - 5p & 2S  - 2P* & 2 - 4  \\ 
Rb XXXVII &	18.225&	6.27E+13 & 1s - 5p & 2S  - 2P* & 2 - 2  \\ 
Rb XXXVII &	17.808&	1.27E+14 & 1s - 4p & 2S  - 2P* & 2 - 4  \\ 
					
Rb XXXVII &	17.797&	1.25E+14 & 1s - 4p & 2S  - 2P* & 2 - 2   \\
Rb XXXVII &	16.896&	3.11E+14 & 1s - 3p & 2S  - 2P* & 2 - 4   \\
Rb XXXVII &	16.870&	3.10E+14 & 1s - 3p & 2S  - 2P* & 2 - 2   \\
Rb XXXVII &	14.299&	1.15E+15 & 1s - 2p & 2S  - 2P* & 2 - 4   \\
Rb XXXVII &	14.209&	1.18E+15 & 1s - 2p & 2S  - 2P* & 2 - 2   \\
					
Rb XXXVI  &	13.891&	1.66E+15 & 1s2 - 1s.2p & 1S  - 1P* & 1 - 3 \\  
Rb XXXVI  &	13.794&	4.58E+14 & 1s2 - 1s.2p & 1S  - 3P* & 1 - 3  \\ 
Ni XXVIII &	10.356&	2.18E+13 & 1s - 5p & 2S  - 2P* & 2 - 4   \\
Ni XXVIII &	10.115&	4.33E+13 & 1s - 4p & 2S  - 2P* & 2 - 4   \\
Ni XXVII  &	9.899 &	3.35E+13 & 1s2 - 1s.5p & 1S  - 1P* & 1 - 3   \\
					
Ni XXVII  &	9.897 &	5.20E+12 & 1s2 - 1s.5p & 1S  - 3P* & 1 - 3   \\
Ni XXVII  &	9.675 &	6.38E+13 & 1s2 - 1s.4p & 1S  - 1P* & 1 - 3   \\
Ni XXVII  &	9.670 &	1.00E+13 & 1s2 - 1s.4p & 1S  - 3P* & 1 - 3   \\
Co XXVII  &	9.622 &	1.88E+13 & 1s - 5p & 2S  - 2P* & 2 - 4 \\  
Ni XXVIII &	9.594 &	1.06E+14 & 1s - 3p & 2S  - 2P* & 2 - 4 \\  
					
Ni XXVIII &	9.585 &	1.06E+14 & 1s - 3p & 2S  - 2P* & 2 - 2 \\  
Co XXVII  &	9.398 &	3.73E+13 & 1s - 4p & 2S  - 2P* & 2 - 4 \\  
Ni XXVII  &	9.191 &	1.63E+14 & 1s2 - 1s.3p & 1S  - 1P* & 1 - 3  \\ 
Co XXVI   &	9.182 &	2.92E+13 & 1s2 - 1s.5p & 1S  - 1P* & 1 - 3 \\  
Co XXVI   &	9.180 &	4.00E+12 & 1s2 - 1s.5p & 1S  - 3P* & 1 - 3  \\ 
					
Ni XXVII  &	9.180 &	2.40E+13 &  1s2 - 1s.3p & 1S  - 3P* & 1 - 3  \\ 
Co XXVI   &	8.975 &	5.68E+13 & 1s2 - 1s.4p & 1S  - 1P* & 1 - 3 \\  
Co XXVI   &	8.971 &	7.70E+12 & 1s2 - 1s.4p & 1S  - 3P* & 1 - 3 \\ 
Fe XXVI   &	8.916 &	1.61E+13 & 1s - 5p & 2S  - 2P* & 2 - 4 \\  
Co XXVII  &	8.914 &	9.17E+13 &  1s - 3p & 2S  - 2P* & 2 - 4  \\ 

\enddata

\end{deluxetable}

\clearpage

\begin{deluxetable}{l c c || c c c }
\tablewidth{0pt}
\tablecaption{Summary of possible explanations for the 13 keV feature of XTE J1810-197 and how each AXP model performs.\label{table:compare}}
\tablehead
{
\multicolumn{1}{c}{Model} & \multicolumn{2}{c}{Cyclotron Emission\tablenotemark{*}} & \multicolumn{3}{c}{Atomic Line Emission}
}

\startdata

\nodata & Proton & Electron & Rubidium & Doppler Broadening & Gravitational Broadening \\[.2 cm]
\cline{2-6}
 &  &  &  &  &  \\[.1 cm] 

Magnetar 	   & Yes & Yes & Unlikely & No  & Unlikely \\[.2cm]
Fall-Back Disk & No  & Yes & Unlikely & No  & No \\[.2cm]
Quark Nova 	   & No  & Yes & Yes	  & Yes & Yes 

\enddata

\tablenotetext{*}{The feasibility of each model under cyclotron emission is for the single case of XTE J1810-197, and does not take into account the argument of multiple sources with the same feature.}

\end{deluxetable}

\end{onecolumn}


\begin{thebibliography}{}

\bibitem[Alford et al.(1999)]{alford_1999} Alford, Mark; Rajagopal, Krishna; Wilczek, Frank.\ 1999, Nuclear Physics B, 537, 443

\bibitem[Boyd \& Sanderson (2003)]{boyd_2003} Boyd, T.J.M., Sanderson, J.J., 2003, The Physics of Plasmas, Cambridge University Press, Cambridge, U.K.; New York, U.S.A.

\bibitem[Charignon et al.(2011)]{char_2011} Charignon, C.; Kostka, M.; Koning, N.; Jaikumar, P.; Ouyed, R.\ 2011, Astronomy \& Astrophysics, 531

\bibitem[Chatterjee et al.(2000)]{chat_2000} Chatterjee, P., Hernquist, L., \& Narayan, R.\ 2000, \apj, 534, 373

\bibitem[Ek{\c s}i \& Alpar (2003)]{eksi_2003} Ek{\c s}i, K. Y., \& Alpar, M. A. \ 2003, ApJ, 599, 450

\bibitem[Ertan \& Alpar (2003)]{ertan_2003} Ertan,U., \& Alpar, M. A.\ 2003, ApJ, 593, L93

\bibitem[Ertan \& Caliskan (2006)]{ertan_2006} Ertan, Ü.; Çalışkan, Ş.\ 2006, ApJ, 649, L87

\bibitem[Ertan et al.(2007)]{ertan_2007} Ertan, {\"U}., Alpar, M.~A., Erkut, M.~H., Ek{\c s}i, K.~Y., \& {\c C}al{\i}{\c s}kan, {\c S}.\ 2007, \apss, 308, 73

\bibitem[Ertan \& Erkut(2008)]{ertan_2008} Ertan, {\"U}., \& Erkut, M.~H.\ 2008, \apj, 673, 1062

\bibitem[Ertan et al. (2009)]{ertan_2009} Ertan, Ü.; Ekşi, K. Y.; Erkut, M. H.; Alpar, M. A.\ 2009, \apj, 702, 1309

\bibitem[Fahlman \& Gregory(1981)]{fahlman_1981} Fahlman, G.~G., \& Gregory, P.~C.\ 1981, \nat, 293, 202

\bibitem[Garcia-Hernandez et al(2006)]{garcia_2006} Garcia-Hernandez, D. A.; Garcia-Lario, P.; Plez, B.; D'Antona, F.; Manchado, A.; Trigo-Rodriguez, J. M.\ 2006, Science, 314, 1751

\bibitem[Gavriil et al.(2008)]{gavriil_2008} Gavriil, F.~P., Dib, R., \& Kaspi, V.~M.\ 2008, 40 Years of Pulsars: Millisecond Pulsars, Magnetars and More, 983, 234

\bibitem[Gotthelf et al.(2004)]{gotthelf_2004} Gotthelf, E.~V., Halpern, J.~P., Buxton, M., \& Bailyn, C.\ 2004, \apj, 605, 368

\bibitem[Ibrahim et al.(2004)]{ibrahim_2004} Ibrahim, A.~I., Markwardt, C.~B., Swank, J.~H., et al.\ 2004, \apjl, 609, L21

\bibitem[Iwazaki(2005)]{iwazaki_2005} Iwazaki, A.\ 2005, \prd, 72, 114003

\bibitem[Jaikumar et al.(2007)]{jaikumar_2007} Jaikumar, P., Meyer, B.~S., Otsuki, K., \& Ouyed, R.\ 2007, \aap, 471, 227

\bibitem[Kwok (2007)]{kwok_2007} Kwok, S.\ 2007, Physics and Chemistry of the Interstellar Medium. University Science Books, 2007.

\bibitem[Lyutikov et al.(2002)]{lyutikov_2002} Lyutikov, M.,  Thompson, C., 
\& Kulkarni, S.~R.\ 2002, Neutron Stars in Supernova Remnants, 271, 262

\bibitem[Niebergal et al.(2006)]{niebergal_2006} Niebergal, Brian; Ouyed, Rachid; Leahy, Denis.\ 2006, \apj, 646, L17

\bibitem[Niebergal et al.(2010)]{niebergal_2010} Niebergal, B., Ouyed, R., \& Jaikumar, P.\ 2010, \prc, 82, 062801

\bibitem[Ouyed et al.(2002)]{ouyed_2002} Ouyed, R., Dey, J., \& Dey, M.\ 2002, \aap, 390, L39

\bibitem[Ouyed et al.(2004)]{ouyed_2004} Ouyed, R., Elgar{\o}y, {\O}., Dahle, H., \& Ker{\"a}nen, P.\ 2004, \aap, 420, 1025

\bibitem[Ouyed et al.(2005)]{ouyed_2005} Ouyed, Rachid; Rapp, Ralf; Vogt, Carsten.\ 2005, \apj, 632, 1001

\bibitem[Ouyed et al.(2006)]{ouyed_2006} Ouyed, R., Niebergal, B., Dobler, W., \& Leahy, D.\ 2006, \apj, 653, 558

\bibitem[Ouyed et al.(2007a)]{ouyed_2007a} Ouyed, R., Leahy, D., \& Niebergal, B.\ 2007, \aap, 473, 357

\bibitem[Ouyed et al.(2007b)]{ouyed_2007b} Ouyed, R., Leahy, D., \& Niebergal, B.\ 2007, \aap, 475, 63

\bibitem[Ouyed \& Leahy(2009)]{ouyed_2009} Ouyed, R., \& Leahy, D.\ 2009, \apj, 696, 562

\bibitem[Pons \& Rea(2012)]{pons_2012} Pons, J.~A., \& Rea, N.\ 2012, \apjl, 750, L6

\bibitem[Ker{\"a}nen et al.(2005)]{keranen_2005} Ker{\"a}nen, P., Ouyed, R., \& Jaikumar, P.\ 2005, \apj, 618, 485

\bibitem[Rea et al.(2008)]{rea_2008} Rea, N., Zane, S., Turolla, R., Lyutikov, M., {\ Gouml}tz, D.\ 2008, \apj, 686, 1245

\bibitem[Sansonetti (2006)]{sansonetti_2006} Sansonetti, J.E.\ 2006, J. Phys. Chem. Ref. Data, v. 35

\bibitem[Steffen et al.(2011)]{steffen_2011} Steffen, W., Koning, N., Wenger, S., Morisset, C., Magnor, M.\ 2011, IEEE Transactions on Visualization and Computer Graphics, 17, 454

\bibitem[Thompson \& Duncan(1995)]{duncan_1995} Thompson, C., \& Duncan, R.~C.\ 1995, \mnras, 275, 255

\bibitem[Vogt et al. (2004)]{vogt_2004} Vogt, C.; Rapp, R.; Ouyed, R.\ 2004, Nuclear Physics A, 375, 543

\bibitem[Wang et al. (2006)]{wang_2006} Wang, Zhongxiang; Chakrabarty, Deepto; Kaplan, David L.\ 2006, Nature, 440, 772

\bibitem[Woods et al.(2005)]{woods_2005} Woods, P.~M., Kouveliotou, C., Gavriil, F.~P., et al.\ 2005, \apj, 629, 985

\bibitem[Woods \& Thompson(2006)]{woods_2006} Woods, P.~M., \& Thompson, C.\ 2006,Compact stellar X-ray sources, 547

\end{thebibliography}
\end{document}